\documentclass[preprints,article,accept,pdftex,moreauthors]{Definitions/mdpi}

\def\Tr{{\rm Tr}}
\def\today{September 1, 2022}	
\firstpage{1} 
\pubvolume{1}
\issuenum{1}
\articlenumber{0}
\pubyear{2022}
\copyrightyear{2022}
\datereceived{} 
\dateaccepted{} 
\datepublished{} 
\hreflink{https://doi.org/} 
\pdfoutput=1

\Title{Numerical interchain mean-field theory for
the specific heat of the bimetallic ferromagnetically coupled chain compound MnNi(NO$_2$)$_4$(en)$_2$ (en = ethylenediamine)}

\TitleCitation{Title}

\Author{Andreas Honecker $^{1}$\orcidA{}*,
Wolfram Brenig $^{2}$,
Maheshwor Tiwari $^{1}$,
Ralf Feyerherm $^{3}$,
Matthias Bleckmann $^{4,5}$, and
Stefan S\"ullow $^{4}$}

\AuthorNames{Andreas Honecker,
Wolfram Brenig,
Maheshwor Tiwari,
Ralf Feyerherm,
Matthias Bleckmann, and
Stefan S\"ullow}

\AuthorCitation{Honecker, A.;
Brenig, W.;
Tiwari, M.;
Feyerherm, R.;
Bleckmann, M.;
S\"ullow, S.}

\address{%
$^{1}$ \quad Laboratoire de Physique Th\'eorique et Mod\'elisation,
  CNRS UMR 8089, CY Cergy Paris Universit\'e, 95302 Cergy-Pontoise Cedex, France\\
$^{2}$ \quad Institut f\"ur Theoretische Physik, TU Braunschweig, 38106 Braunschweig,
Germany \\
$^{3}$ \quad Helmholtz-Zentrum Berlin f\"ur Materialien und Energie GmbH, 14109 Berlin,
Germany \\
$^{4}$ \quad Institut f\"ur Physik der Kondensierten Materie, TU Braunschweig, 38106 Braunschweig, Germany \\
$^{5}$ \quad Wehrwissenschaftliches Institut f\"ur Werk- und Betriebsstoffe (WIWeB), 85435 Erding, Germany
}

\corres{Correspondence: andreas.honecker@cyu.fr}

\abstract{We present a detailed study of the field-dependent specific 
heat of the bimetallic ferromagnetically coupled chain compound 
MnNi(NO$_2$)$_4$(en)$_2$, en = ethylenediamine. For this material, which 
in zero field orders antiferromagnetically below $T_N=2.45$~K, small 
fields suppress magnetic order. Instead, in such fields a double-peak 
like structure in the temperature dependence of the specific heat is 
observed. We attribute this behavior to the existence of an acoustic and 
an optical mode in the spin wave dispersion as result of the existence of 
two different spins per unit cell. We compare our experimental data to 
numerical results for the specific heat obtained by exact diagonalization 
and Quantum-Monte-Carlo simulations for the alternating spin chain model, 
using parameters that have been derived from the high-temperature 
behavior of the magnetic susceptibility. The interchain coupling is 
included in the numerical treatment at the mean-field level. We observe 
remarkable agreement between experiment and theory, including the 
ordering transition, using previously determined parameters. Furthermore, 
the observed strong effect of an applied magnetic field on the ordered 
state of MnNi(NO$_2$)$_4$(en)$_2$ promises interesting magnetocaloric 
properties.}

\keyword{quantum spin chains;
specific heat;
Quantum Monte Carlo simulations;
exact diagonalization;
mean-field theory}

\begin{document}

\section{Introduction}

Alternation in spin systems, be it of the magnetic coupling, the local 
symmetry or the spin value, induces new and exotic types of magnetic 
ground states and excitations 
\cite{Mikeska2004,Dender97,Oshikawa97,Essler98,zvyagin,Tiegel16,hagiwara,wolfram,bartolome,drillon,Pati1997a,Pati1997b,Ivanov2000,Kolezhuk1997,Yamamoto1998a,Yamamoto1998b,Nakanishi2002,Yamamoto2005}. 
In particular, this is exemplified in novel bimetallic chain systems, {\it 
viz.}, molecule-based chain systems with alternately arranged magnetic 
units carrying quantum spins $S_1$ and $S_2$ of different size. The 
ability to synthesize mixed-spin chain materials 
\cite{bartolome,kahn,Caneschi89,Zhou94,Nishizawa2000,Yao12,MENG2019134,Thorarinsdottir20,Yamaguchi21} 
has stimulated theoretical investigations 
\cite{drillon,Pati1997a,Pati1997b,Ivanov2000,Kolezhuk1997,Yamamoto1998a,Yamamoto1998b,Nakanishi2002,Yamamoto2005,fukushima,FUKUSHIMA20051409,PhysRevB.73.014411,PhysRevB.76.224410,Yuan_2010,HU2015539,Yan15,Silva2021}.
The magnon dispersion relation of such chains splits into an optical and an acoustical mode because of the two differently sized quantum spins $S_1$ and $S_2$ per unit cell, both for antiferromagnetic coupling along the chain \cite{drillon,Pati1997a,Pati1997b,Ivanov2000,Kolezhuk1997,Yamamoto1998a,Yamamoto1998b,Nakanishi2002,Silva2021} as well as ferromagnetic coupling \cite{Yamamoto2005,fukushima,FUKUSHIMA20051409,Yuan_2010,HU2015539,Yan15}. Although ground state and fundamental excitations of a Heisenberg ferromagnet are simple, thermodynamic properties are very sensitive to interactions of the magnon excitations, as is evidenced by the ferromagnetic uniform spin-1/2 Heisenberg chain (compare chapter 11.3 of \cite{takahashi99} and references therein). Computations for bimetallic Heisenberg chains show that the two energy scales associated  to the acoustic and the optical spin excitation modes are reflected by ``double-peak'' kind of features in the specific heat $c_p(T)$ for both antiferro- \cite{drillon,Pati1997a,Pati1997b,Ivanov2000,Kolezhuk1997,Yamamoto1998a,Yamamoto1998b,Nakanishi2002} and ferromagnetic \cite{Yamamoto2005,fukushima,FUKUSHIMA20051409} coupling.

Similar predictions had been made for chains of mixed classical and quantum spins as far back as 1975 \cite{Dembiski75}.
In spite of this long history, experimental verifications of the features expected in the specific heat are lacking. In fact, experimental tests of mixed-spin chain models are scarce \cite{Hagiwara1998,Hagiwara1999,Fujiwara2000}, since most materials available contain
elements with larger spins \cite{Caneschi89,Zhou94,Nishizawa2000,wynn,affronte99,Girtu00,Lascialfari03},
which are difficult to be treated adequately in theoretical calculations \cite{drillon,Pati1997a,Pati1997b,Ivanov2000,Kolezhuk1997,Yamamoto1998a,Yamamoto1998b,Nakanishi2002}.

Here we will present a verification of the two energy scale prediction 
via a detailed study of the specific heat $c_p$ of 
MnNi(NO$_2$)$_4$(en)$_2$, en = ethylenediamine = C$_2$N$_2$H$_4$, in zero 
and applied fields. After the field-induced suppression of long-range 
antiferromagnetic order we observe a double-peak like structure in the 
temperature dependence of $c_p$ for MnNi(NO$_2$)$_4$(en)$_2$. We compare 
our findings with the results of numerical calculations for an $S_1 = 1$, 
$S_2 = 5/2$ mixed spin chain in zero and external fields. We demonstrate 
that the in-field calculations, for which finite size effects are 
negligible, fully reproduce the double-peak structure of the 
experimentally observed in-field specific heat. This shows that the optic 
and acoustic spin excitation mode are reflected by the thermodynamics of 
this bimetallic chain system. Quantum Monte Carlo (QMC) simulations of 
the individual chains augmented by a self-consistent mean-field treatment 
of interchain coupling even yields a remarkably accurate description of 
the ordering transition in a vanishing magnetic field.

The remainder of this manuscript is organized as follows: Section 
\ref{sec:Exp} presents some more details on MnNi(NO$_2$)$_4$(en)$_2$ and 
in particular a measurement of its specific heat. We then proceed in 
Sec.~\ref{sec:Theory} with a detailed theoretical analysis based on exact 
diagonalization and QMC simulations combined with a mean-field treatment 
of interchain coupling; some complementary details are provided in 
Appendix~\ref{sec:MFTsc}. In Sec.~\ref{sec:MCE} we briefly comment on 
magnetocaloric properties of MnNi(NO$_2$)$_4$(en)$_2$ before we summarize 
our findings in Sec.~\ref{sec:Concl}. The Appendix~\ref{sec:MFT} contains 
a summary of a complementary single-site mean-field treatment.

\section{Experiment}

\label{sec:Exp}

\begin{figure}[t!]
\begin{center}
\includegraphics[width=0.8\columnwidth]{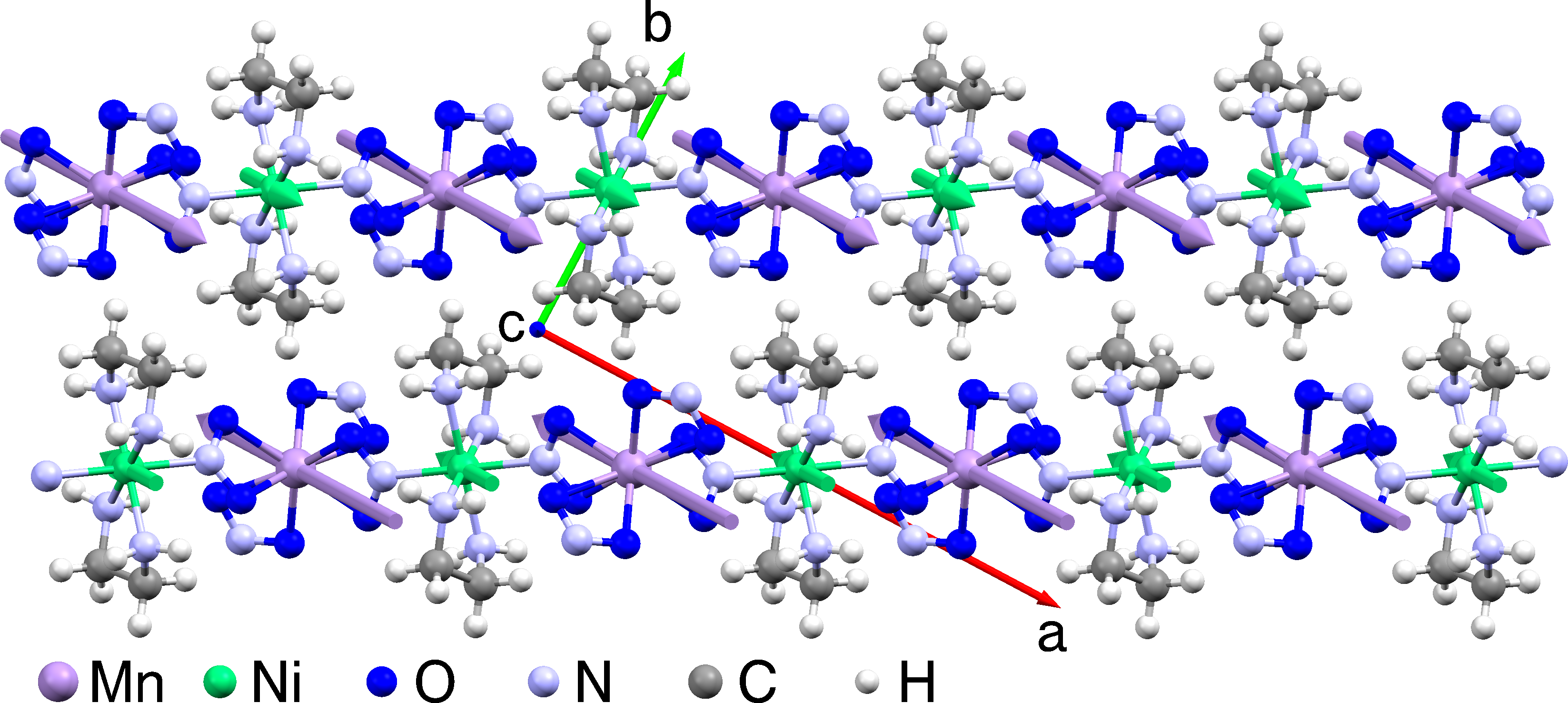}
\end{center}
\caption{Two chains in the $a$-$b$ plane of MnNi(NO$_2$)$_4$(en)$_2$, based
on the crystal structure of Ref.~\cite{Gillon2002}. The thick arrows on the
Mn and Ni atoms show the zero-field ordered state: ferromagnetic along the
chains and antiferromagnetic between chains.
}
\label{fig:fig1}
\end{figure}

\subsection{MnNi(NO$_2$)$_4$(en)$_2$ (en = ethylenediamine)}

MnNi(NO$_2$)$_4$(en)$_2$ is one of the best characterized mixed spin 
chain compounds \cite{Kahn97,feyerherm,Gillon2002,KREITLOW20052413}, 
crystallizing in an orthorhombic structure, space group {\it Pccn} 
(lattice parameters $a = 14.675$~\AA, $b = 7.774$~\AA, $c = 12.401$~\AA). 
It contains chains of alternately arranged Ni and Mn ions linked by 
NO$_2$ ligands, which carry magnetic moments with spin $S_1 = 1$ and $S_2 
= 5/2$, respectively (Fig.~\ref{fig:fig1}). The magnetic coupling along 
the chain, $J$, is ferromagnetic \cite{feyerherm}, with $J = 2.8$~K 
\cite{fukushima}. A finite ionic zero-field splitting $D$ of 0.36~K is 
derived from the anisotropy of the susceptibility. Because of an 
effective antiferromagnetic interchain coupling of $J_{\perp} = 0.036$~K, 
the system undergoes a transition into an antiferromagnetically (AFM) 
ordered state below $T_N = 2.45$~K in zero magnetic field and at ambient 
pressure \cite{fukushima,feyerherm,KREITLOW20052413}. The long-range 
magnetically ordered state is suppressed by rather small magnetic fields 
\cite{feyerherm}.

\subsection{Specific heat}

\label{sec:SpecHeat}

For our study we have used crystals MnNi(NO$_2$)$_4$(en)$_2$ investigated 
previously \cite{feyerherm}.
Here, we will present the easy axis data $B \| c$, for which AFM ordering is suppressed in $\sim 0.3$~T.
The heat capacity was measured using commercial calorimeters in magnetic fields $B \| c$ up to 1.6~T at temperatures $T$ down to 0.4~K. As will be discussed below, these $c$ axis data allow a comparison to more accurate numerical calculations than the data $\| a$.

\begin{figure}[t!]
\begin{center}
\includegraphics[width=0.6\columnwidth]{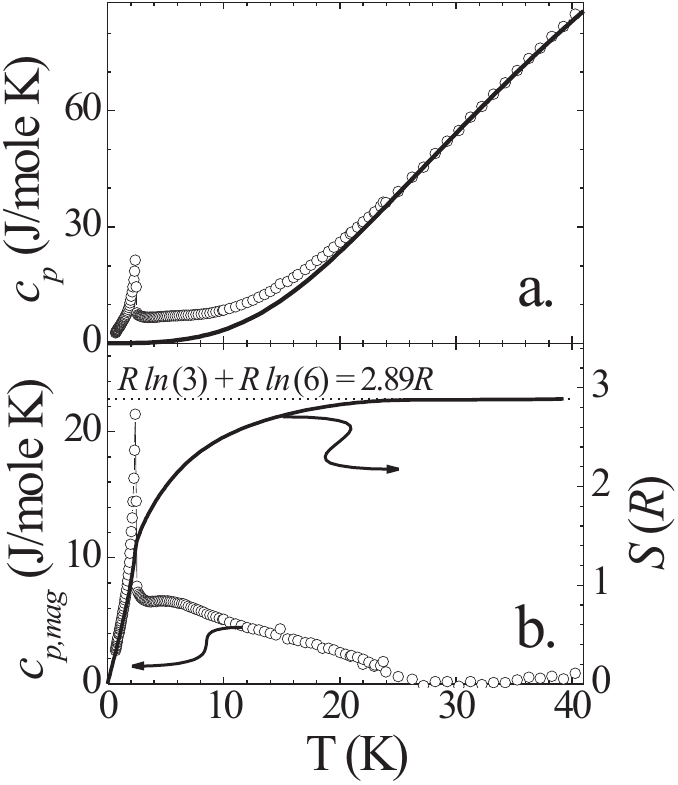}
\end{center}
\caption{(a) Zero-field specific heat $c_p$ of MnNi(NO$_2$)$_4$(en)$_2$ 
as function of temperature $T$. (b) Zero-field magnetic specific heat 
$c_{p,mag}$ and associated entropy $S$ per mole of 
MnNi(NO$_2$)$_4$(en)$_2$ as function of temperature.}
\label{fig:fig2}
\end{figure}

In Fig.~\ref{fig:fig2}(a) we depict the zero-field specific heat $c_p$ of 
MnNi(NO$_2$)$_4$(en)$_2$ as function of $T$. The AFM anomaly at $T_N = 
2.45$~K is clearly discernible. To derive the magnetic specific heat we 
determine the lattice contribution $c_{p,lat}$. Since a single $T^3$-term 
does not reproduce the experimental data above $T_N$, we use two Debye 
contributions, each calculated via the full Debye-integral, to 
parameterize $c_{p,lat}$. MnNi(NO$_2$)$_4$(en)$_2$ is built up by chain 
segments -Mn-NO$_2$-Ni-NO$_2$-, with two ethylendiamine molecules and two 
NO$_2$ groups attached to the Mn and Ni ions, respectively 
(Fig.~\ref{fig:fig1}). Intra-molecular oscillations of ethylendiamine or 
NO$_2$, because of the light atoms involved, yield Einstein 
contributions, which are irrelevant for the temperatures considered here. 
The chain segment units Mn, Ni and NO$_2$ are similar in atomic weight. 
Therefore, to parameterize the lattice contribution of these units we 
choose one Debye-temperature $\Theta_D$ with $3 \times 4 = 12$ modes. 
Analogously, the four attached molecules ethylendiamine and NO$_2$ per 
chain segment are parameterized by a second Debye-temperature 
contributing with 12 modes. This way, we reproduce the lattice specific 
heat of MnNi(NO$_2$)$_4$(en)$_2$ with Debye-temperatures $\Theta_{D1} = 
138$~K and $\Theta_{D2} = 249$~K (solid line in Fig.~\ref{fig:fig2}(a)).

We obtain the magnetic specific heat contribution $c_{p,mag}$ by 
subtracting $c_{p,lat}$ from the total $c_p$ (Fig.~\ref{fig:fig2}(b)). 
Further, by numerically integrating $c_{p,mag}/T$ we obtain the magnetic 
entropy $S$ included in Fig.~\ref{fig:fig2}(b). Both quantities indicate 
that above $T_N$ there are magnetic fluctuations present over a wide 
temperature range. In $c_{p,mag}$ there is a broad anomaly ranging up to 
$\sim 10\,T_N$. The associated entropy reaches only $1.4\,R$ at 
$T_N$, which is less than half of the value expected for the sum of the 
magnetic entropies of Ni ($S_1 = 1$) and Mn ($S_2 = 5/2$), $R \ln(3) + R 
\ln(6) \approx 2.89\,R$ (dotted line in Fig.~\ref{fig:fig2}(b)). This 
value is reached only at 10\,$T_N$. Note that the saturation of $S$ at 
$2.89\,R$ demonstrates the consistency and adequacy of our derivation of 
the lattice specific heat.

\begin{figure}[t!]
\begin{center}
\includegraphics[width=0.6\columnwidth]{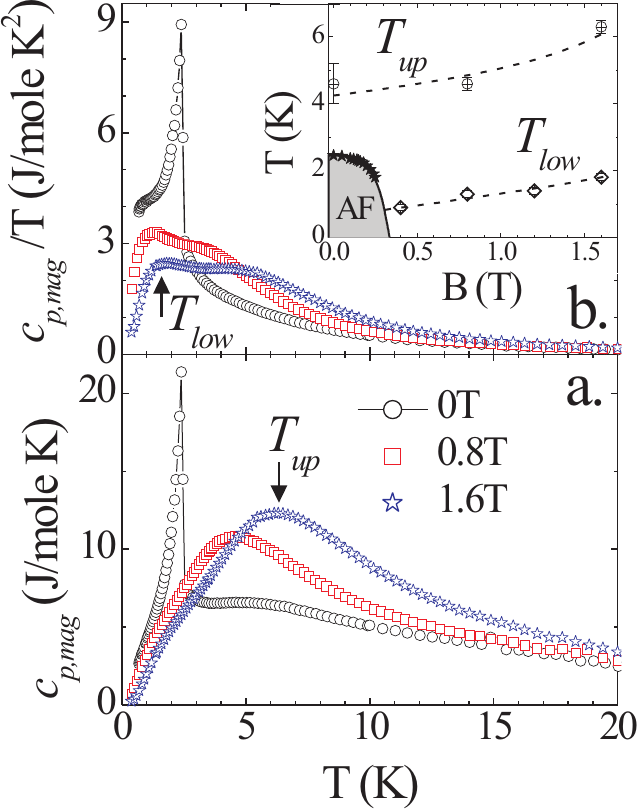}
\end{center}
\caption{(a) Field dependence of $c_{p,mag}$ of MnNi(NO$_2$)$_4$(en)$_2$ for fields $B \| c$.
(b) The same data  plotted as $c_{p,mag}/T$. Inset: The magnetic phase diagram of  MnNi(NO$_2$)$_4$(en)$_2$ for $B \| c$: $T_N$ from Ref.~\cite{feyerherm} ($\star$), $T_{\rm up}$ from the maximum in $c_{p,mag}$ ($\oplus$), $T_{\rm low}$ from the maximum in $c_{p,mag}/T$ ($\diamond$); lines are guides to the eye.} \label{fig:fig3}
\end{figure}

AFM order in MnNi(NO$_2$)$_4$(en)$_2$ is suppressed by small magnetic 
fields \cite{feyerherm}. This enables us to study magnetic fluctuations 
in MnNi(NO$_2$)$_4$(en)$_2$, as they appear in $c_p$. In 
Fig.~\ref{fig:fig3} we plot $c_{p,mag}$ as function of field. We observe 
a rapid suppression of the AFM state, in agreement with 
Ref.~\cite{feyerherm}. Moreover, after suppression of the AFM state the 
broad specific heat anomaly above $T_N$ becomes much more pronounced in 
magnetic fields, and is clearly visible already in the non-phonon 
corrected data.

The temperature $T_{\rm up}$ of the maximum in $c_{p,mag}$ represents a 
measure for an energy scale characteristic for the magnetic fluctuation 
spectrum (indicated for the 1.6~T data in Fig.~\ref{fig:fig3}). In the 
inset of Fig.~\ref{fig:fig3} we record its field dependence up to 1.6~T, 
with a modest increase of $T_{\rm up}$ by about 1~K/T. Further, after 
suppression of AFM order in the $T$ dependence of $c_p$ there is 
additional structure. This is most clearly seen for $c_{p,mag}/T$, where 
one now observes a double-peak like structure (see Fig.~\ref{fig:fig3}(b)). 
We take as measure for a second characteristic energy scale $T_{\rm low}$ 
the maximum in $c_{p,mag}/T$ and include its field dependence in 
Fig.~\ref{fig:fig3}. Again, we find a modest increase of $T_{\rm low}$ by 
about 1~K between 0.4 and 1.6~T.

$T_{\rm up}$ and $T_{\rm low}$ are clearly distinct temperatures and 
increase at a similar rate. Therefore, they do not stem from ionic states 
Zeeman split in an external field. Further, extrapolating $T_{\rm low}$ 
to zero field yields a finite value of about 0.7~K, implying that $T_{\rm 
low}$ does not arise from Zeeman splitting of ionic degenerate states. 
Therefore, we associate both characteristic energy scales $T_{\rm up}$ 
and $T_{\rm low}$ with collective excitation modes of the magnetic 
fluctuation spectrum of MnNi(NO$_2$)$_4$(en)$_2$ as result of the 
existence of an acoustic and an optical magnon mode.

\section{Theory}

\label{sec:Theory}

We now proceed to provide a theoretical description of the experimental 
findings.

\subsection{Model}

We start from the basic chain model
\begin{equation}
H = -J \sum_{x=1}^{N/2} \left(\vec{S}_x \cdot \vec{s}_x +
\vec{s}_x \cdot \vec{S}_{x+1}\right) 
- D \sum_{x=1}^{N/2} \left(S^z_x\right)^2 - h \sum_{x=1}^{N/2}
\left(S^z_x + s^z_x\right) \, ,
\label{eq:Hop}
\end{equation}
where the $\vec{s}_x$ ($\vec{S}_x$) correspond to the spins of the Ni 
ions (Mn ions) and have $S_1 = 1$ ($S_2 = 5/2$). Following 
Refs.~\cite{fukushima,feyerherm}, we take a single-ion anisotropy into 
account only for the Mn sites. The main role of this anisotropy is to 
select a preferred axis, it should not matter too much if this is due to 
the Mn or the Ni sites, and it is the form Eq.~(\ref{eq:Hop}) for which 
parameters were extracted in Ref.~\cite{fukushima} by analyzing the 
high-temperature behavior of the magnetic susceptibility. Nevertheless, 
we refer to Appendix~\ref{sec:1mag} for a discussion of the one-magnon 
dispersion for the case where both anisotropies are present. In the 
following discussion we will use the parameters that have been determined 
in Ref.~\cite{fukushima}, namely $J = 2.8$~K and $D=0.36$~K, or in units 
with $J=1$: $D=0.36/2.8 \approx 0.129$. In the latter units, and assuming 
magnetic $g$ factors $g=2$, the magnetic fields of $0.8$~T and $1.6$~T 
shown in Fig.~\ref{fig:fig3} are modeled by $h=0.4$ and $0.8$, 
respectively.

\subsection{Numerical treatment of decoupled chains}

\label{sec:TheoryDecoupled}

\begin{figure}[t!]
\begin{center}
\includegraphics[width=0.49\columnwidth]{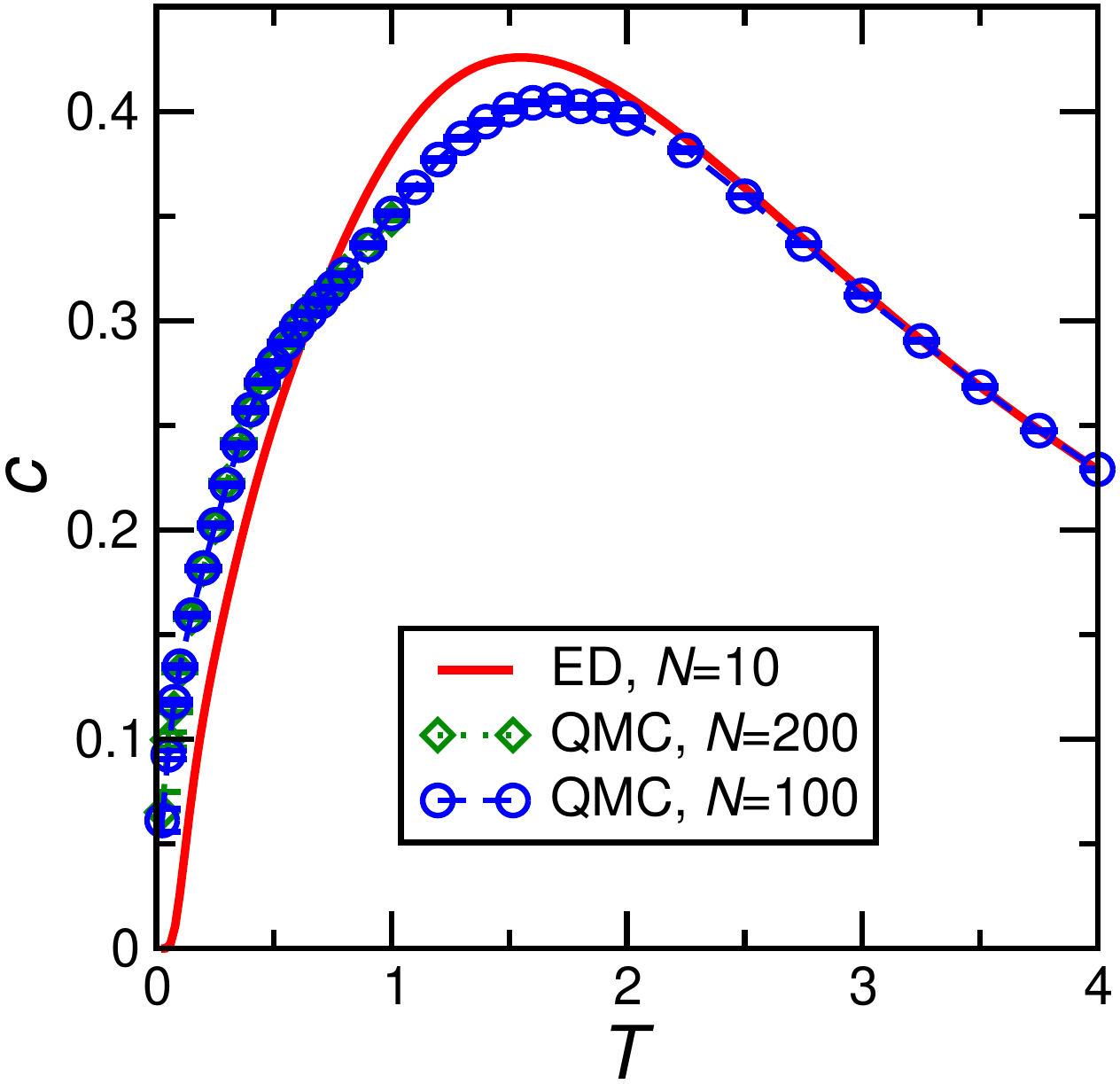}\hfill
\includegraphics[width=0.49\columnwidth]{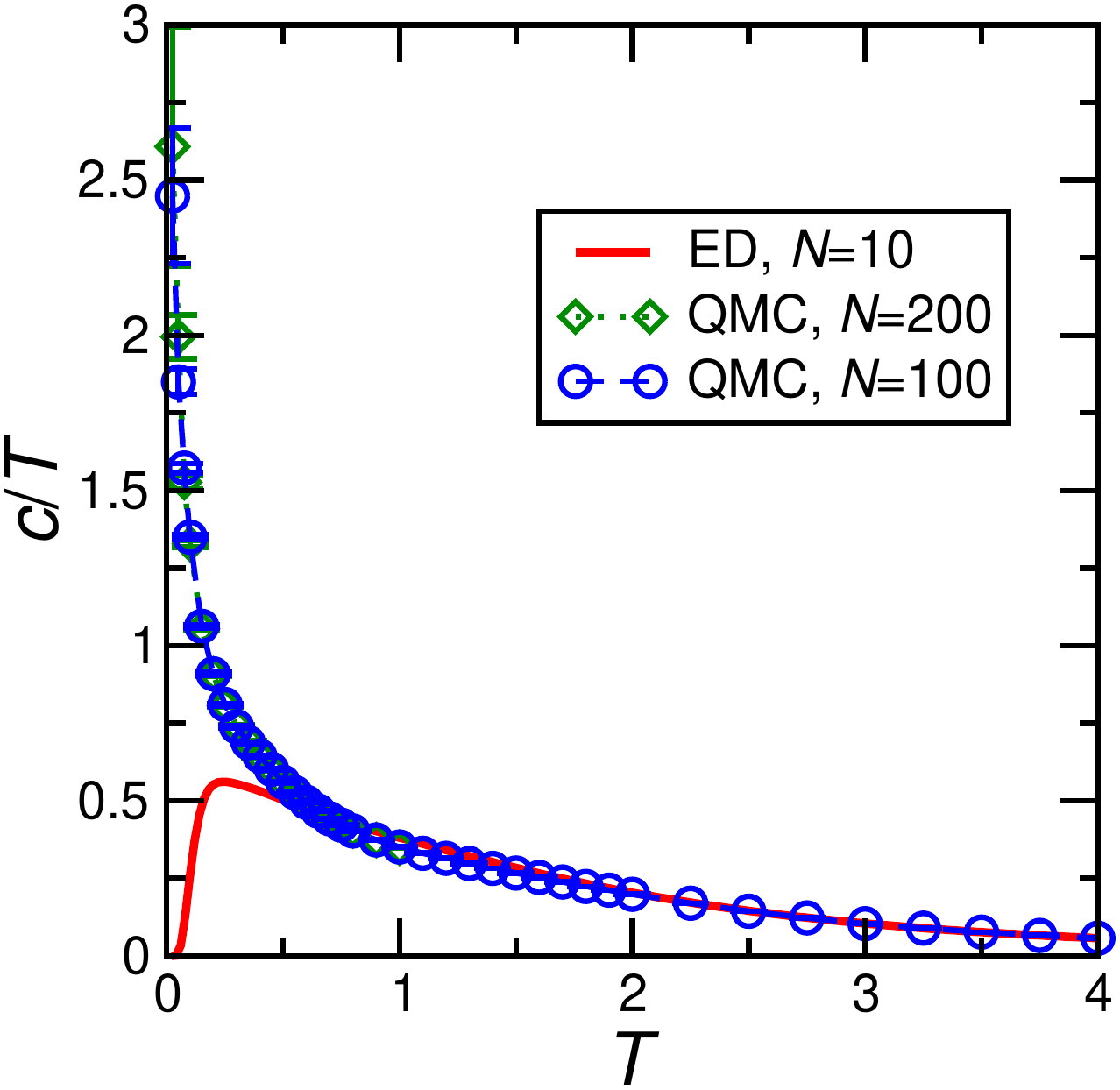}
\end{center}
\vspace*{-3mm}
\caption{\label{fig:Ch0D0}
Specific heat per spin $c$ calculated for $D=0$, $h=0$, $J=1$ by exact 
diagonalization (ED) and Quantum Monte Carlo (QMC). The left panel shows 
the specific heat itself while the right panel shows the specific heat 
divided by temperature $c/T$.
}
\end{figure}

Previously, some of the present authors have performed exact (full) 
diagonalization and Quantum Monte Carlo (QMC) simulations of chains with 
$S_1=1/2$, $S_2=1$ \cite{fukushima,FUKUSHIMA20051409}. The previous exact 
diagonalization (ED) investigations went to $N=14$ spins with $S_1=1/2$ 
and $S_2=1$. When we replace a spin $1/2$ by $5/2$, the local Hilbert 
space dimension increases from 2 to 6, {\it i.e.}, by a factor 3. Thus, 
here we have to contend ourselves with ED for chains with $N=10$. Adding 
one unit cell would increase the total Hilbert space dimension by a 
factor 18 for the case $S_1=1$, $S_2=5/2$ such that next system size 
$N=12$ remains out of reach. We use conservation of $S^z$ as well as 
spatial symmetries. The magnetic susceptibility $\chi$ and specific heat 
$c$ can then be calculated from the eigenvalues and the associated 
quantum numbers.

To access longer chains, we use QMC. The present QMC simulations have 
been carried out with the ALPS \cite{alps1,alps2} directed loop 
applications \cite{alps-sse,PhysRevLett.87.047203} in the stochastic 
series expansion framework \cite{Sandvik}. To be precise, these 
computations were started a while ago. We have therefore used version 1.3 
of the ALPS applications \cite{alps1} rather than the more recent release 
2.0 \cite{Bauer_2011}. The specific heat in a magnetic field can be 
sensitive to the pseudorandom-number generator such that this needs to be 
carefully chosen. Here we have used the ``Mersenne Twister 19937'' (MT) 
pseudorandom-number generator \cite{MTrng}. To verify reliability of our 
results, we have performed QMC simulations for $N=10$ (data not shown 
here) and double-checked them against our ED computations for the same 
system size.

\begin{figure}[t!]
\begin{center}
\includegraphics[width=0.49\columnwidth]{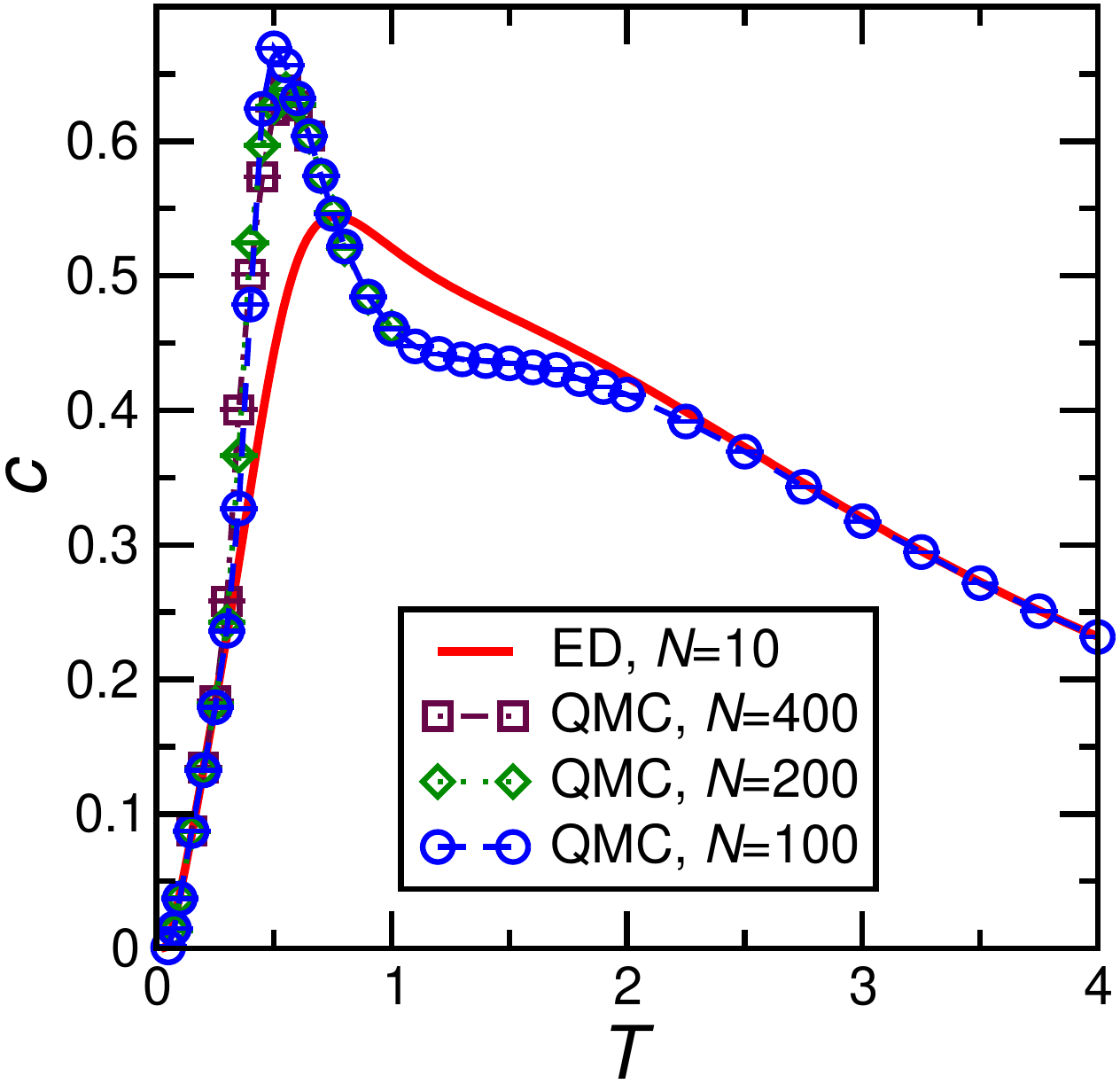}\hfill
\includegraphics[width=0.49\columnwidth]{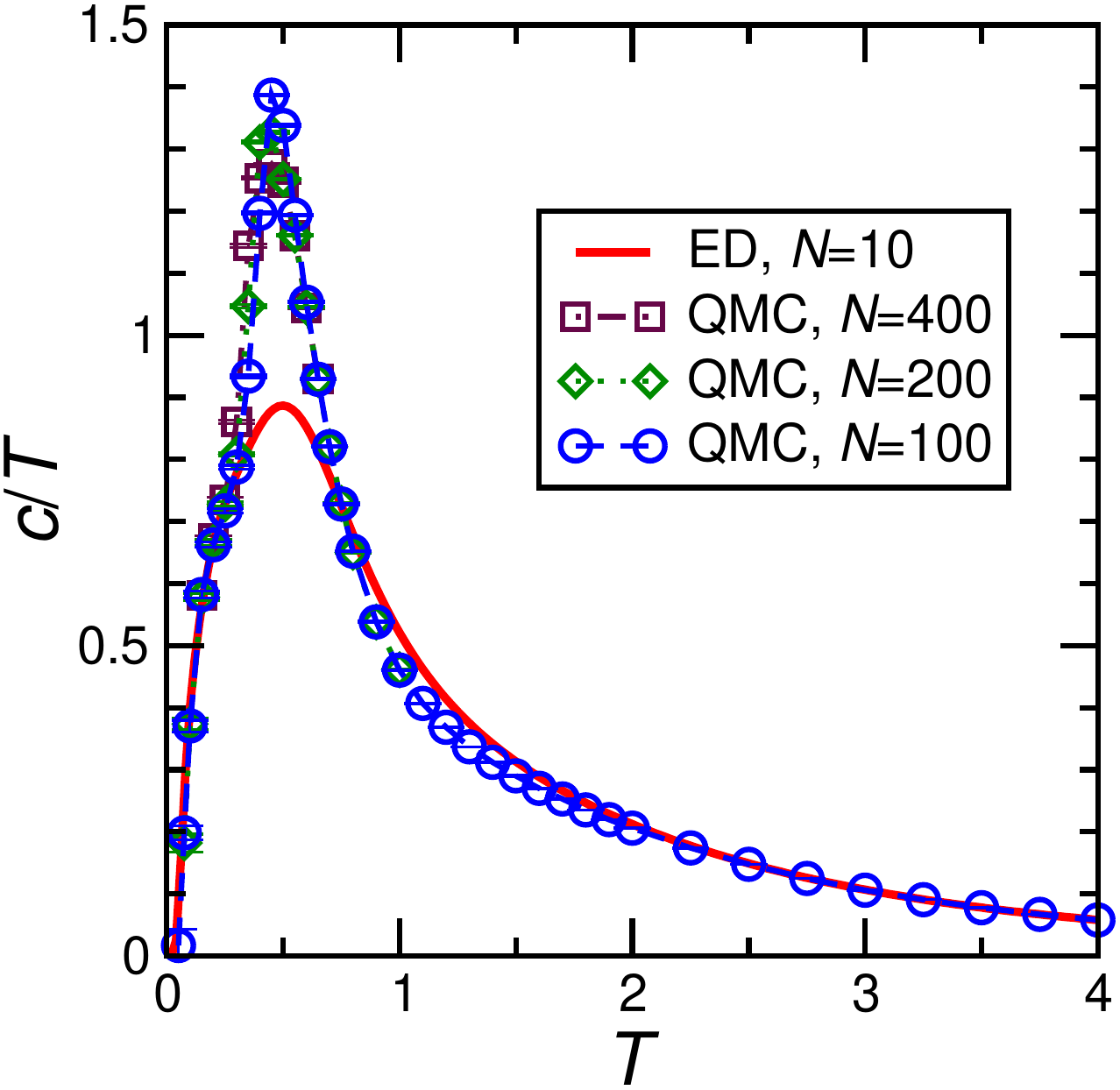}
\end{center}
\vspace*{-3mm}
\caption{\label{fig:Ch0Df}
Specific heat per spin $c$ calculated for $D=0.36/2.8$, $h=0$, $J=1$.
As in Fig.~\ref{fig:Ch0D0}, the left panel
shows the specific heat itself while the right panel shows
the specific heat divided by temperature $c/T$.
}
\end{figure}

\begin{figure}[t!]
\begin{center}
\includegraphics[width=0.69\columnwidth]{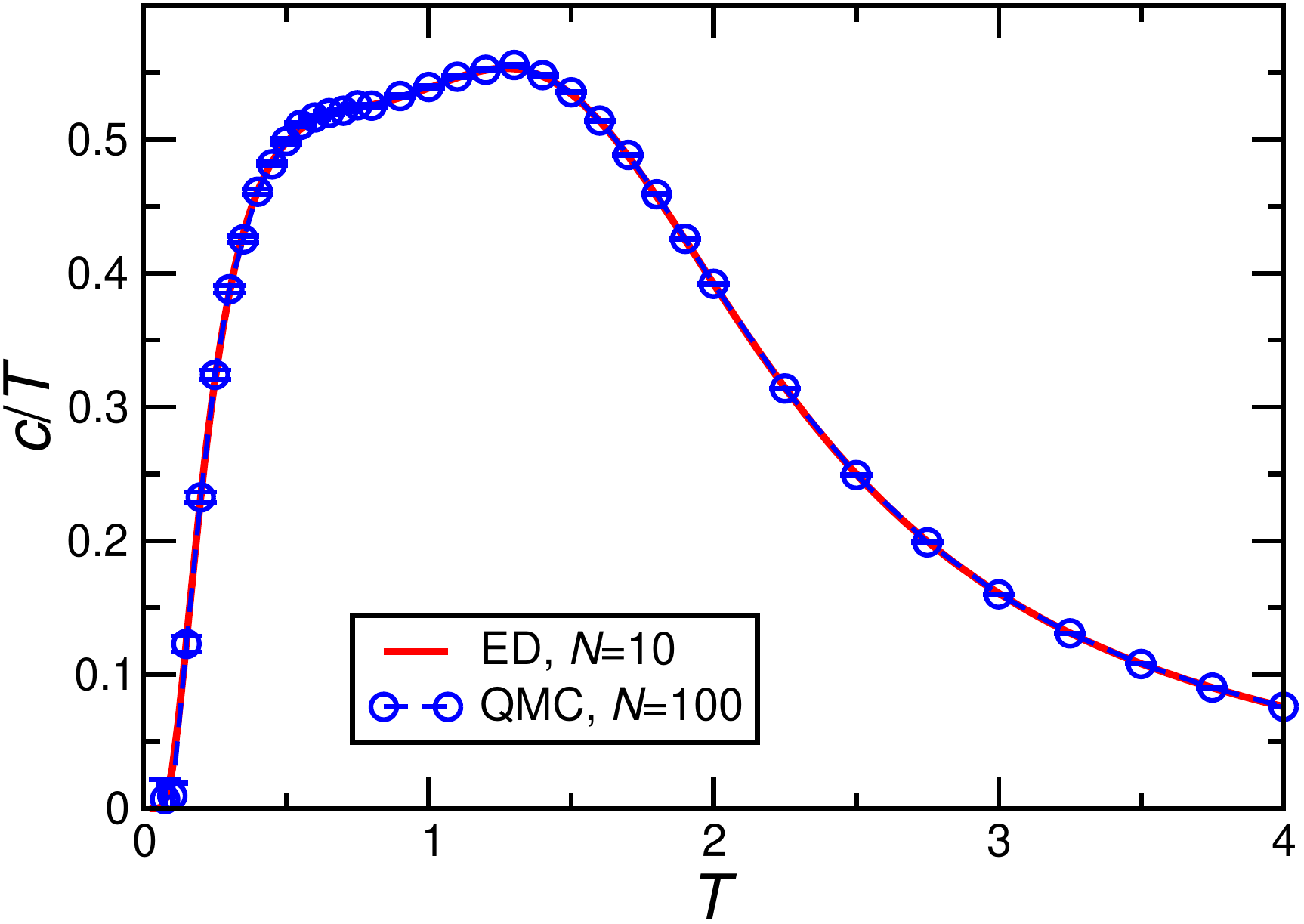}
\end{center}
\vspace*{-3mm}
\caption{\label{fig:Ch0.4Df}
Specific heat per spin divided by temperature $c/T$ calculated for 
$D=0.36/2.8$, $J=1$ in a magnetic field $h=0.4$.
}
\end{figure}

\begin{figure}[t!]
\begin{center}
\includegraphics[width=0.69\columnwidth]{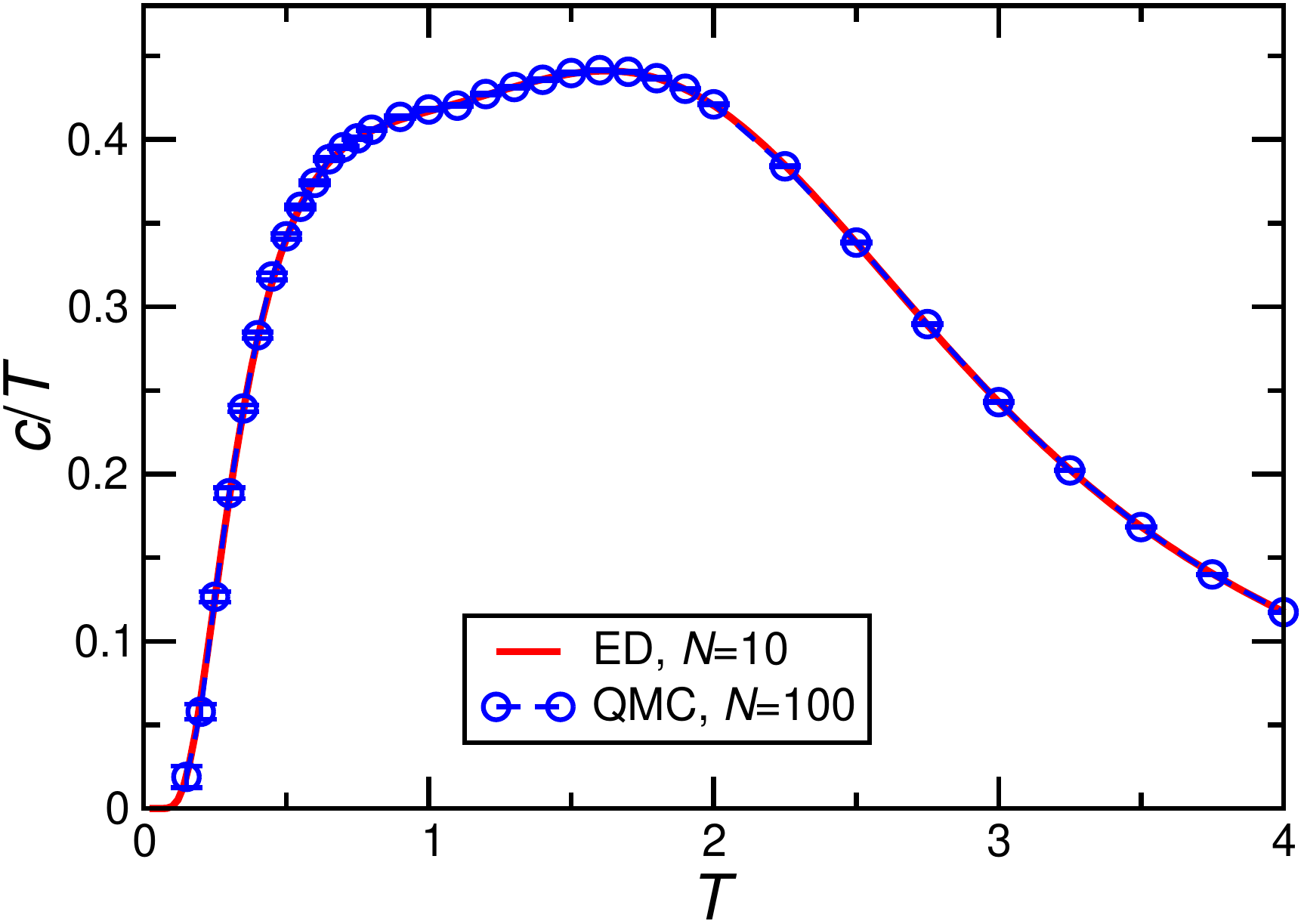}
\end{center}
\vspace*{-3mm}
\caption{\label{fig:Ch0.8Df}
Specific heat per spin divided by temperature $c/T$ calculated
for $D=0.36/2.8$, $J=1$ in a magnetic field $h=0.8$.
}
\end{figure}

Figures \ref{fig:Ch0D0}--\ref{fig:Ch0.8Df} show ED ($N=10$) and QMC ($N 
\ge 100$) results for the specific heat. The QMC simulations become 
challenging at the lowest temperatures, in particular for finite $D$ and 
$h$. This leads to visible statistical error bars at low $T$ in 
particular in Figs.~\ref{fig:Ch0.4Df} and \ref{fig:Ch0.8Df} while 
otherwise statistical errors are negligible. For $h=0$, finite-size 
effects are relevant, as demonstrated by visible deviations between the 
$N=10$ and $100$ data in Figs.~\ref{fig:Ch0D0} and \ref{fig:Ch0Df}. On 
the other hand, no further change is visible for larger $N$, {\it i.e.}, 
$N=100$ can be considered as representative of the thermodynamic limit 
for $h=0$. Finally, a field of $h \ge 0.4\,J$ lifts the ground-state 
degeneracy and opens a sufficiently large gap in the spectrum such that 
$N=10$ and $N=100$ become indistinguishable (see Figs.~\ref{fig:Ch0.4Df} 
and \ref{fig:Ch0.8Df}) and $N=10$ ED suffices to describe the 
thermodynamic limit.

For $h=0$ and $D=0$, the ground state is an $SU(2)$ multiplet with 
$(7\,N/2+1)$ components. This leads to a difference between the 
zero-temperature entropies per site for $N=10$ and $N=100$ of $\Delta 
S=0.299744\ldots$. Accordingly, the entropy integral $\int_{0}^{\infty} 
{\rm d}T \, c/T$, {\it i.e.}, the corresponding area under the $N=100$ 
curve of the right panel of Fig.~\ref{fig:Ch0D0} is expected to be bigger 
than of the corresponding $N=10$ curve by this amount $\Delta S$. The QMC 
data for the specific heat $c$ not only exhibits a maximum at $T \approx 
1.8\,J$, but also a shoulder at $T \approx 0.5\,J$ (see left panel of 
Fig.~\ref{fig:Ch0D0}), corresponding to the two expected features 
\cite{fukushima,FUKUSHIMA20051409}.

Figure \ref{fig:Ch0Df} shows the result with the single-ion anisotropy 
$D>0$ included, still at $h=0$. The presence of the single-ion anisotropy 
reduces the ground-state degeneracy to two and opens a gap in the 
one-magnon spectrum, see Appendix \ref{sec:1mag} for details. For $N=10$, 
the resulting ground-state entropy $\ln 2$ is still almost 5\%\ of the 
total entropy. This leads to a difference between the zero-temperature 
entropies per site for $N=10$ and $N=100$ of $\Delta S = 
\ln2/10-\ln2/100=0.062383\ldots$. While this is smaller than in the case 
of $D=0$, the difference is still visible in the ED data as compared to 
QMC, compare the right panel of Fig.~\ref{fig:Ch0Df}. From the point of 
view of physics, the specific heat $c$ in the left panel of 
Fig.~\ref{fig:Ch0Df} may be more instructive. The shoulder-like feature 
for $D=0$ has developed into a sharp peak around $T \approx 0.5\,J$ for 
the value $D=0.36/2.8$ while in turn the previous global maximum of $c$ 
has become a shoulder around $T\approx 1.7\,J$. In any case, these two 
features can be traced from $D=0$ to finite $D$.

Finally, we add a magnetic field $h>0$, corresponding to the experimental 
case where we actually observed two features in the specific heat (see 
Fig.~\ref{fig:fig3}(b)). Application of a finite field $h>0$ not only 
lifts the remaining ground-state degeneracy, but $h \ge 0.4\,J$ opens a 
sufficiently large gap in the spectrum such that finite-size effects are 
negligible already for $N=10$, as mentioned before and shown in 
Figs.~\ref{fig:Ch0.4Df} and \ref{fig:Ch0.8Df}. Like in the experiment, we 
observe the emergence of a double-peak structure where both the feature 
at $T\approx 0.5\,J$ and in particular the one at $T/J = 1.5 \ldots 2$ 
shifts to higher temperatures with increasing magnetic field, compare 
Figs.~\ref{fig:Ch0.4Df} and \ref{fig:Ch0.8Df}.

\subsection{Mean-field treatment of the interchain coupling}

\label{sec:ChainMFT}

In zero external magnetic field, an antiferromagnetic phase transition 
with a N\'eel temperature $T_N=2.45\,{\rm K}=0.875\,J$ is observed 
experimentally, as discussed in Sec.~\ref{sec:Exp}. This demonstrates 
that interchain coupling should be included into a quantitative 
description, at least for $h=0$ and $T \lesssim J$, even if the numerical 
results of Sec.~\ref{sec:TheoryDecoupled} already qualitatively reproduce 
the experiment in a finite magnetic field.

Since the chains are ferromagnetic, we assume that only the total magnetization of one chain acts via an effective field on the neighboring chains.
Thus, we start from a family of interchain mean-field Hamiltonians\footnote{To be precise,
one starts from a coupling between chains $i$ and $j$ of the form
\begin{equation}
\frac{J_{i,j}}{2\,N}
\sum_{x=1}^{N/2} \left(\vec{S}_{i,x} + \vec{s}_{i,x}\right)
\cdot
\sum_{y=1}^{N/2} \left(\vec{S}_{j,y} + \vec{s}_{j,y}\right)
\label{eq:Hop1}
\end{equation}
which one replaces by
\begin{equation}
\frac{J_{i,j}}{2} \, \langle M_i\rangle
\sum_{y=1}^{N/2} \left(S^z_{j,y} + s^z_{j,y}\right)
+ \frac{J_{i,j}}{2} \, \langle M_j\rangle
\sum_{x=1}^{N/2} \left(S^z_{i,x} + s^z_{i,x}\right)
- N \, \frac{J_{i,j}}{2} \, \langle M_i\rangle \, \langle M_j\rangle \, .
\end{equation}
We drop the term $- N \, \frac{J_{i,j}}{2} \, \langle M_i\rangle \, \langle M_j\rangle$
for the time being, but one should remember to add
this term for total energy computations and in particular if one wants to
write expectation values as derivatives of the free energy, see also Ref.~\cite{GZZ16}.
}
\begin{equation}
H_i^{\rm MF} = -J \sum_{x=1}^{N/2} \left(\vec{S}_x \cdot \vec{s}_x +
\vec{s}_x \cdot \vec{S}_{x+1}\right) 
- D \sum_{x=1}^{N/2} \left(S^z_x\right)^2
 - \left(h-\sum_{j\ne i} J_{i,j}\,\langle M_j\rangle\right) \sum_{x=1}^{N/2}
\left(S^z_x + s^z_x\right) \, ,
\label{eq:HopMFi}
\end{equation}
where the magnetization of the $i$th chain should satisfy the self-consistency condition
\begin{equation}
N\,\langle M_i\rangle
= \frac{\Tr \left(\sum\limits_{x=1}^{N/2} \left(S^z_x + s^z_x\right)\,
{\rm e}^{-\beta\,H_i^{\rm MF}}\right)}
{\Tr \left({\rm e}^{-\beta\,H_i^{\rm MF}}\right)}
=  \frac{\Tr \left(\sum\limits_{x=1}^{N/2} \left(S^z_x + s^z_x\right)\,
{\rm e}^{-\beta\,\left(H_i^{\rm MF}- N \, \sum\limits_{j\ne i} \frac{J_{i,j}}{2} \, \langle M_i\rangle \, \langle M_j\rangle\right)}\right)}
{\Tr \left({\rm e}^{-\beta\,\left(H_i^{\rm MF}- N \, \sum\limits_{j\ne i} \frac{J_{i,j}}{2} \, \langle M_i\rangle \, \langle M_j\rangle\right)}\right)}
\label{eq:MiMF}
\end{equation}
with $\beta = 1/T$ ($k_B=1$), as usual. The assumption of only average 
magnetizations of one chain affecting the neighboring ones is motivated by 
the exact exchange paths between chains in MnNi(NO$_2$)$_4$(en)$_2$ being 
unknown (compare the crystal structure of Fig.~\ref{fig:fig1}), and was 
also made in Ref.~\cite{fukushima}.

We now 
consider two cases.
Firstly, for $h=0$, we expect antiferromagnetic order that
should be described by two types of chains $i=1,2$. Furthermore,
by symmetry one expects that $\langle M_1\rangle = -\langle M_2\rangle = \langle M\rangle$. This sign difference
can be absorbed by a spin inversion on every other chain, which
also flips the sign of the interchain coupling. Therefore,
we introduce an effective interchain coupling
$J_\perp  = -\sum_{j\ne i} J_{i,j}$, where the minus sign will allow
us to treat all chains
as having the same magnetization $\langle M\rangle \ge 0$.
Secondly, for $h \ge 0.4$, one stays in a paramagnetic phase where we expect all chain magnetizations to be equal
$\langle M_i\rangle = \langle M\rangle$. Now we straightforwardly set the effective interchain coupling $J_\perp  = \sum_{j\ne i} J_{i,j}$.

Under either of these assumptions, the family of mean-field Hamiltonians
(\ref{eq:HopMFi}) reduces to a single interchain mean-field Hamiltonian
\begin{equation}
H^{\rm MF} = H_{1D}
 - \left(h-J_\perp\,\langle M\rangle\right) \, N\,M \, ,
\label{eq:HopMF}
\end{equation}
with
\begin{equation}
H_{1D} =  -J \sum_{x=1}^{N/2} \left(\vec{S}_x \cdot \vec{s}_x +
\vec{s}_x \cdot \vec{S}_{x+1}\right) 
- D \sum_{x=1}^{N/2} \left(S^z_x\right)^2 \, , \qquad
N\,M = \sum_{x=1}^{N/2}
\left(S^z_x + s^z_x\right) \, .
\label{eq:DefOpsMF}
\end{equation}
The magnetization should now satisfy the modified
self-consistency condition
\begin{equation}
\langle M\rangle
= \frac{\Tr \left(M\,
{\rm e}^{-\beta\,H^{\rm MF}}\right)}
{\Tr \left({\rm e}^{-\beta\,H^{\rm MF}}\right)}
= \frac{\Tr \left(M\,
{\rm e}^{-\beta\,\left(H^{\rm MF}-N\,\frac{J_\perp}{2} \langle M \rangle^2\right)}\right)}
{\Tr \left({\rm e}^{-\beta\,\left(H^{\rm MF}-N\,\frac{J_\perp}{2} \langle M \rangle^2\right)}\right)}
\, .
\label{eq:MMF}
\end{equation}
Recall that in order to cast both the antiferromagnetic case at $h=0$ and the paramagnetic
case at $h>0$ in the same single-chain form, it was necessary
to introduce different signs for the effective interchain coupling $J_\perp$
in the two cases. Still, the absolute value of $J_\perp$ is the same in both cases.

\begin{figure}[tb!]
\begin{center}
\includegraphics[width=0.69\columnwidth]{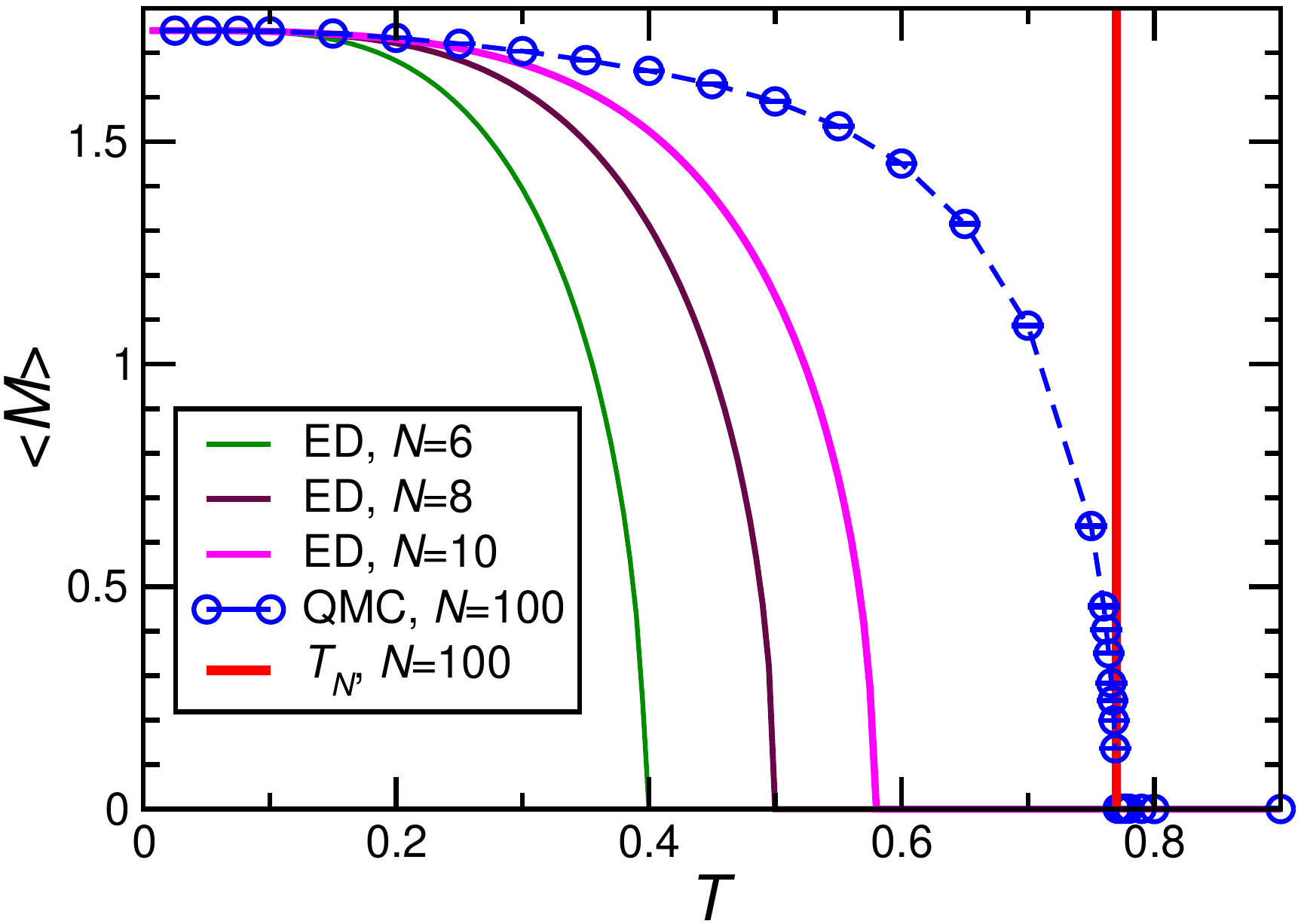}
\end{center}
\vspace*{-3mm}
\caption{\label{fig:magMFT}
Self-consistent mean-field results for the magnetization $\langle 
M\rangle$ with $J_\perp = -0.072/2.8$, $D=0.36/2.8$, $h=0$, $J=1$. The 
vertical line indicates the estimate $T_N = 0.77$ for the N\'eel 
temperature obtained from chains with $N=100$ sites (for details see 
text).
}
\end{figure}

Since the magnetization $\langle M\rangle$ is easily evaluated even
within QMC, it is possible to run a self-consistency loop
using a numerical evaluation of the chain magnetization $\langle M\rangle$,
{\it i.e.}, one starts with an initial guess such as $\langle M\rangle = 7/4$,
recomputes $\langle M \rangle$ from Eq.~(\ref{eq:MMF}) and iterates until a
desired level of accuracy is reached, compare Appendix~\ref{sec:MFTsc} for some further details.
Some ED and QMC results for the self-consistent $\langle M\rangle$
for $h=0$ are shown in Fig.~\ref{fig:magMFT}.
The vertical line in Fig.~\ref{fig:magMFT} shows an estimate of the 
N\'eel temperature that will be discussed in the following subsection 
\ref{sec:Chi}. One observes in Fig.~\ref{fig:magMFT} that the estimated 
N\'eel temperature $T_N$ varies by almost a factor two as one goes from 
$N=6$ to $100$ spins in a chain. Even with $N=10$, one still deviates by 
about 25\%\ from the estimate obtained with $N=100$. On the other hand, 
analysis of the data shown in Figs.\ \ref{fig:Ch0Df} and \ref{fig:chiTN} 
below indicates that $N=100$ should indeed be sufficient to represent the 
thermodynamic limit along the chains.

\subsubsection{Magnetic susceptibility and ordering temperature}

\label{sec:Chi}

The numerical treatment of a single chain yields direct access to
\begin{equation}
\chi_{1D} = \beta\,N\,\left(\langle M^2 \rangle - \langle M \rangle^2\right) \, ,
\label{eq:defChiMF1}
\end{equation}
where $\langle M\rangle$ may be included in the self-consistent effective
field, but is considered to be {\em fixed}, {\it i.e.}, contributions
from the self-consistent field are \emph{not} included in Eq.~(\ref{eq:defChiMF1}).

The magnetic susceptibility should be defined by
\begin{equation}
\chi_{\rm MF} = \frac{\partial}{\partial h}\,\langle M\rangle
\label{eq:defChiMF2}
\end{equation}
within the interchain mean-field approximation.
Insertion of the definition Eq.~(\ref{eq:MMF}) for the magnetization
and some straightforward algebra leads to
\begin{equation}
\chi_{\rm MF} = \left(1-J_\perp\,\chi_{\rm MF}\right) \, \chi_{1D} \, .
\label{eq:chiMF}
\end{equation}
The result (\ref{eq:chiMF}) can be solved  for $\chi_{\rm MF}$
and one finds\footnote{As a consequence of the spin inversion that we have applied to half of the chains at $h=0$, this is actually not the uniform, but a staggered susceptibility in the case of a vanishing external field.
}
\begin{equation}
\chi_{\rm MF} = \frac{\chi_{1D}}{1+J_\perp\,\chi_{1D}} \, .
\label{eq:chiRPA}
\end{equation}
This approximation is widely used in the literature
(see for example \cite{Schulz,Cavadini2000,TodoShibasaki})
and also known under the name ``random-phase-approximation''.
Since there are some similarities with the Stoner model of
ferromagnetism (see, e.g., chapter 7.4 of \cite{Fazekas}), one
can also call $1+J_\perp\,\chi_{1D}$ a ``Stoner factor''.
Note that the above derivation is essentially the same as the
computation on page 66 of \cite{GrDiplom}, but the linearizing
assumption $\langle M\rangle \approx h\,\chi_{\rm MF}$ has
been dropped. Accordingly, we see that Eq.~(\ref{eq:chiRPA}) also
applies for a finite magnetization $\langle M\rangle \ne 0$ of a
single chain.


\begin{figure}[t!]
\begin{center}
\includegraphics[width=0.69\columnwidth]{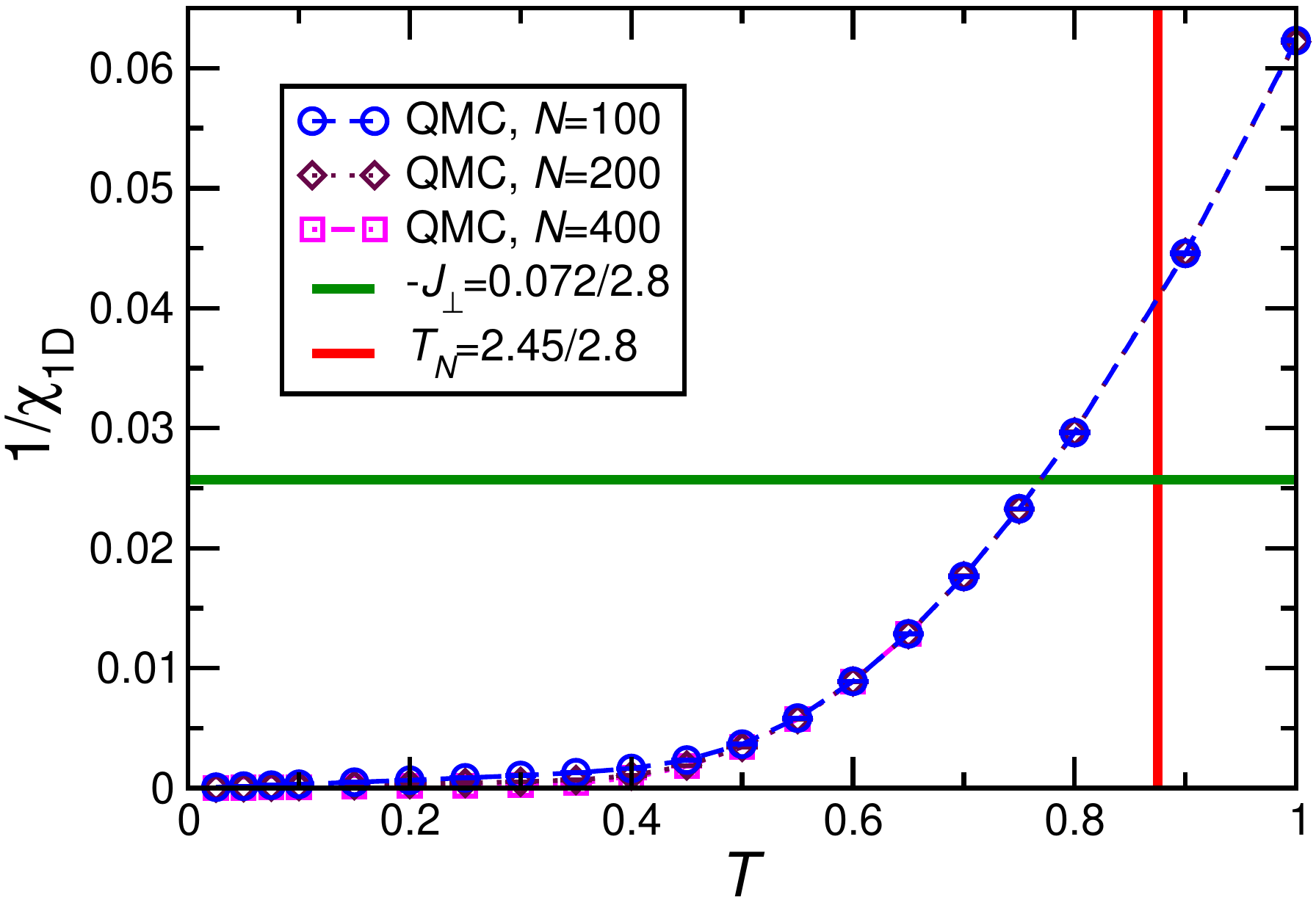}
\end{center}
\vspace*{-3mm}
\caption{\label{fig:chiTN}
Inverse magnetic susceptibility calculated by QMC for a single chain
with $D=0.36/2.8$, $h=0$, $J=1$.
Also shown are the estimated value of the
interchain coupling  $-J_\perp = 0.072$~K
\cite{fukushima} as well as the experimental N\'eel temperature
$T_N = 2.45$~K \cite{fukushima,feyerherm} divided by $J=2.8$~K.
}
\end{figure}

A zero of the denominator in Eq.~(\ref{eq:chiRPA}) signals a second-order
phase transition. This yields the standard condition for the
N\'eel temperature
\begin{equation}
-J_\perp = \frac{1}{\chi_{1D}(T_N)} \, .
\label{eq:chiTN}
\end{equation}
Let us use this condition to take a look at the ordering transition in
zero external field where $\langle M \rangle = 0$ for $~T > T_N$ such
that Eq.~(\ref{eq:chiTN}) can be evaluated without running a self-consistency
cycle.
Our QMC results for $1/\chi_{1D}$ at $h=0$
are shown in Fig.~\ref{fig:chiTN}
for $N=100$, $200$, and $400$. The fact that these three system sizes are essentially indistinguishable on the scale of the figure shows that $N=100$ suffices to represent the thermodynamic limit $N=\infty$.

If one assumes the value $-J_\perp = 0.072$~K (horizontal line in 
Fig.~\ref{fig:chiTN}) that has been deduced in \cite{fukushima} by fitting 
the magnetic susceptibility for $T \ge 10$~K, one reads off an ordering 
temperature
$T_N \approx 0.77\,J \approx 2.16$~K. This deviates by about 12\%
from the experimental value $T_N=2.45$~K, which is remarkably good for a mean-field
theory. Conversely, if one insists on the experimental
value $T_N = 0.875\,J$, one infers an interchain coupling
$-J_\perp \approx 0.04\,J \equiv 0.11$~K which is about 50\%\
larger than the estimate of \cite{fukushima}. In fact,
$1/\chi_{1D}$ varies quite strongly in this temperature range.
Therefore, $T_N$ is not very sensitive to the interchain coupling
$J_\perp$.

In any case, an interchain coupling of a few percent suffices to yield an 
antiferromagnetic ordering temperature at $h=0$ that is of the same order 
as the coupling in an individual chain, reflecting strong ferromagnetic 
ordering tendencies of the decoupled chains.

\subsubsection{Specific heat}

Let us now take a closer look at the specific heat in interchain mean-field theory.
As in the case of the magnetic susceptibility, the numerical treatment of the individual chains provides convenient access to
\begin{equation}
c_{1D} 
= \frac{\beta^2}{N}\,\left(\langle \left(H^{\rm MF}\right)^2 \rangle - \langle H^{\rm MF} \rangle^2\right) \, ,
\label{eq:defCMF1}
\end{equation}
where $\langle M\rangle$ may again be included in the self-consistent effective
field, but is considered to be {\em fixed}.

The self-consistent magnetization $\langle M\rangle$ is also 
temperature-dependent such that the
specific heat should be written as a first derivative of the internal energy
\begin{equation}
c_{\rm MF}
=\frac{1}{N}\, \frac{\partial\,U}{\partial T}\,
=\frac{1}{N}\,
\frac{\partial}{\partial T}\,\left(\langle H^{\rm MF}\rangle-N\,\frac{J_\perp}{2} \langle M \rangle^2\right)
\, .
\label{eq:ddefCMF2}
\end{equation}
The temperature derivative can in principle be calculated 
numerically. For reasons of numerical stability, in particular in a Monte-Carlo setting, it is nevertheless preferable to carry the derivatives out analytically. Since we are not aware of such an analysis having been presented before, we present it here in some detail.
With the help of $[M,H^{\rm MF}]=0$, we find from Eq.~(\ref{eq:ddefCMF2}) that
\begin{eqnarray}
c_{\rm MF} &=& -\frac{\beta^2}{N}\,
\frac{\partial}{\partial \beta}\,\left(\langle H^{\rm MF}\rangle-N\,\frac{J_\perp}{2} \langle M \rangle^2\right)
 \nonumber \\
&=& c_{1D}+\beta^3\,J_\perp \, \frac{\partial \,\langle M\rangle}{\partial \beta}\,
\left( \langle H^{\rm MF} \, M \rangle
-  \langle H^{\rm MF} \rangle \langle M \rangle \right) \, .
\label{eq:defCMF2}
\end{eqnarray}
This expression contains another derivative $\frac{\partial \,\langle M\rangle}{\partial \beta}$
for which we can find an expression that is very similar to Eq.~(\ref{eq:chiRPA})
(including a ``Stoner factor'' $1+J_\perp\,\chi_{1D}$):
\begin{equation}
\frac{\partial \,\langle M\rangle}{\partial \beta}
 = -\frac{\langle H^{\rm MF} \, M \rangle - 
   \langle H^{\rm MF} \rangle \langle M \rangle}{1+J_\perp\,\chi_{1D}} \, .
\label{eq:dMdBeta}
\end{equation}
%
Noting the relation
\begin{equation}
\left. \frac{\partial \,\langle M\rangle}{\partial T}\right|_{h,1D}
 = \beta^2 \, \left( \langle H^{\rm MF} \, M \rangle - 
   \langle H^{\rm MF} \rangle \langle M \rangle \right) \, ,
\label{eq:dMdT}
\end{equation}
the combination of Eqs.~(\ref{eq:defCMF2}) and (\ref{eq:dMdBeta})
can also be written in the following form:
\begin{equation}
c_{\rm MF} =  c_{1D}-
\frac{J_\perp}{\beta} \frac{1}{1+J_\perp\,\chi_{1D}} \,
\left(\left. \frac{\partial \,\langle M\rangle}{\partial T}\right|_{h,1D}\right)^2
\label{eq:CMF2final}
\end{equation}
In this form, the sign of the second term is evident. This form is also 
useful for the purpose of evaluation since Eq.~(\ref{eq:CMF2final}) 
contains only quantities that can be related to static expectation values 
for a single chain with a {\em fixed} value of $\langle M\rangle$ via 
Eqs.~(\ref{eq:defChiMF1}), (\ref{eq:defCMF1}), and (\ref{eq:dMdT}). The 
only object that is non-standard is the crosscorrelator in 
Eq.~(\ref{eq:dMdT}), but it represents exactly the same observable as was 
used in Ref.~\cite{mceTrippe} to compute the adiabatic cooling rate by 
QMC.

\subsubsection{Comparison with experimental specific heat}

\label{sec:SecCompExp}

\begin{figure}[tb!]
\begin{center}
\includegraphics[width=0.69\columnwidth]{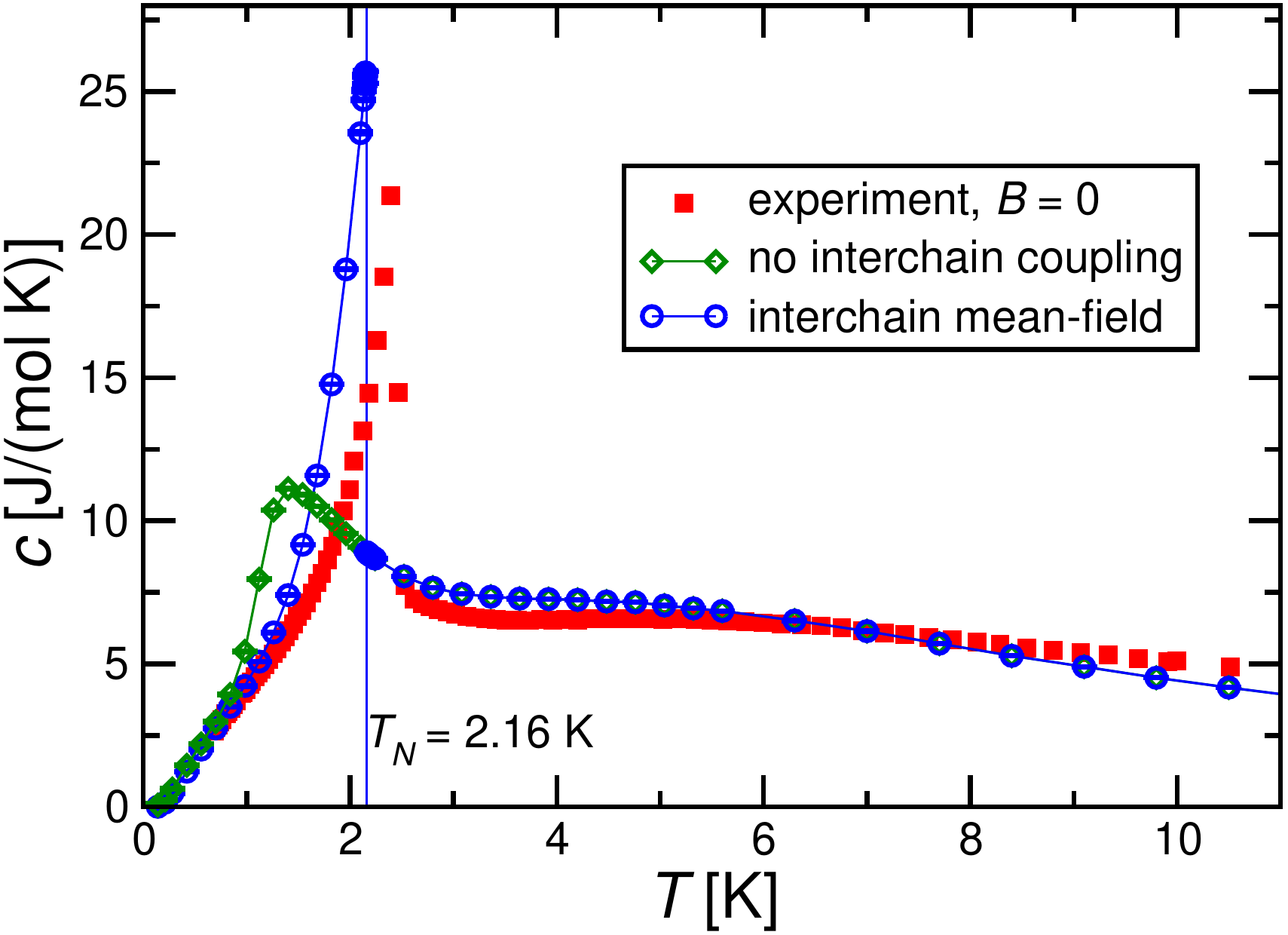}
\end{center}
\vspace*{-3mm}
\caption{\label{fig:CMFTh0}
Magnetic specific heat per spin $c$ for $J=2.8$~K, $D=0.36$~K in zero 
magnetic field $h=0$ in comparison with the experimental results for 
$B=0$. Theoretical results are obtained by QMC with $N=100$ both for 
decoupled chains ($J_\perp=0$) and with a self-consistent mean-field 
treatment for $J_\perp = -0.072$~K.
}
\end{figure}

\begin{figure}[tb!]
\begin{center}
\includegraphics[width=0.69\columnwidth]{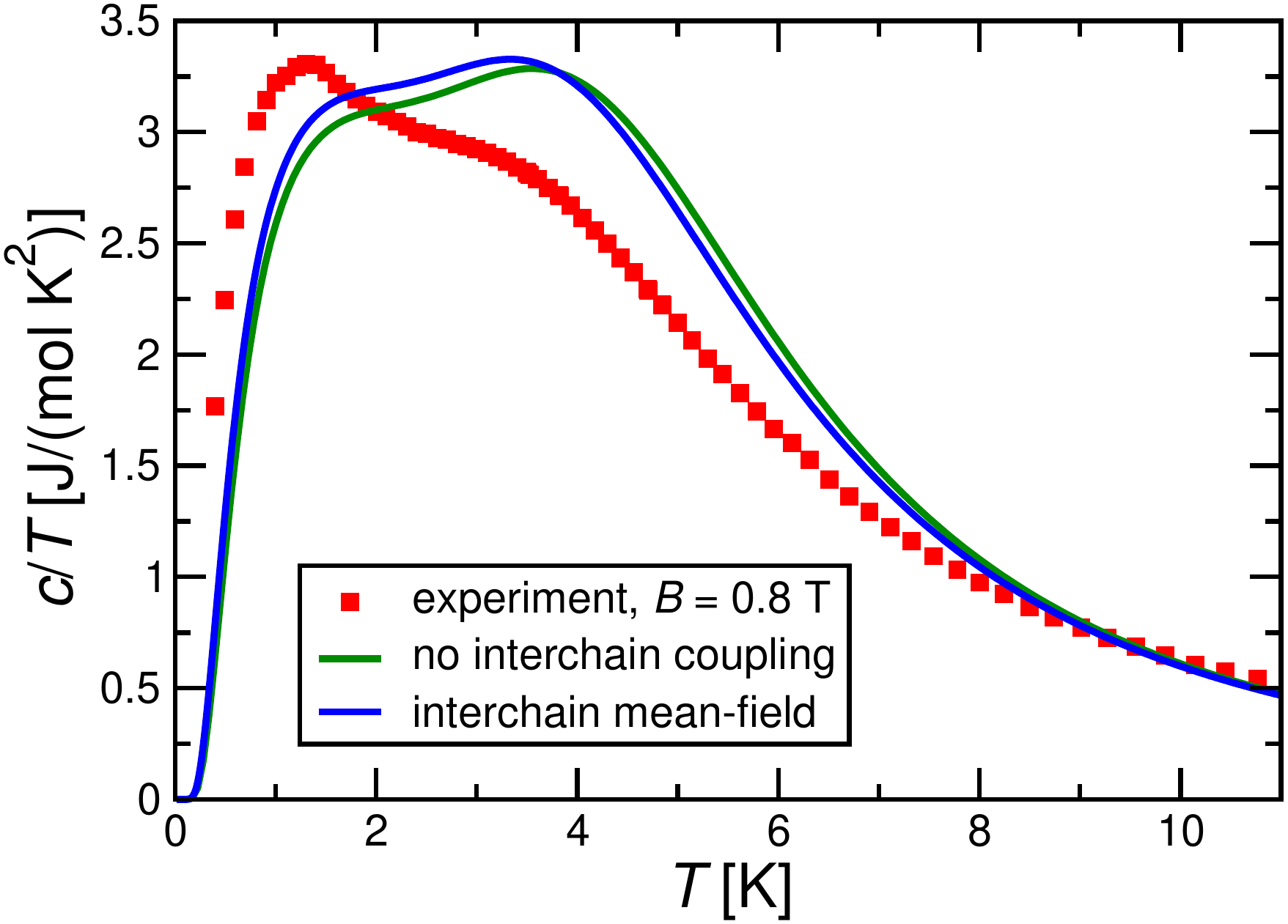}
\end{center}
\vspace*{-3mm}
\caption{\label{fig:CMFTh0.4}
Magnetic specific heat per spin divided by temperature $c/T$ for 
$J=2.8$~K, $D=0.36$~K in a magnetic field $h=0.4\,J$ in comparison with 
the experimental results for $B=0.8$~T. Theoretical results are obtained 
by ED with $N=10$ both for decoupled chains ($J_\perp=0$) and with a 
self-consistent mean-field treatment for $J_\perp = 0.072$~K.
}
\end{figure}

\begin{figure}[tb!]
\begin{center}
\includegraphics[width=0.69\columnwidth]{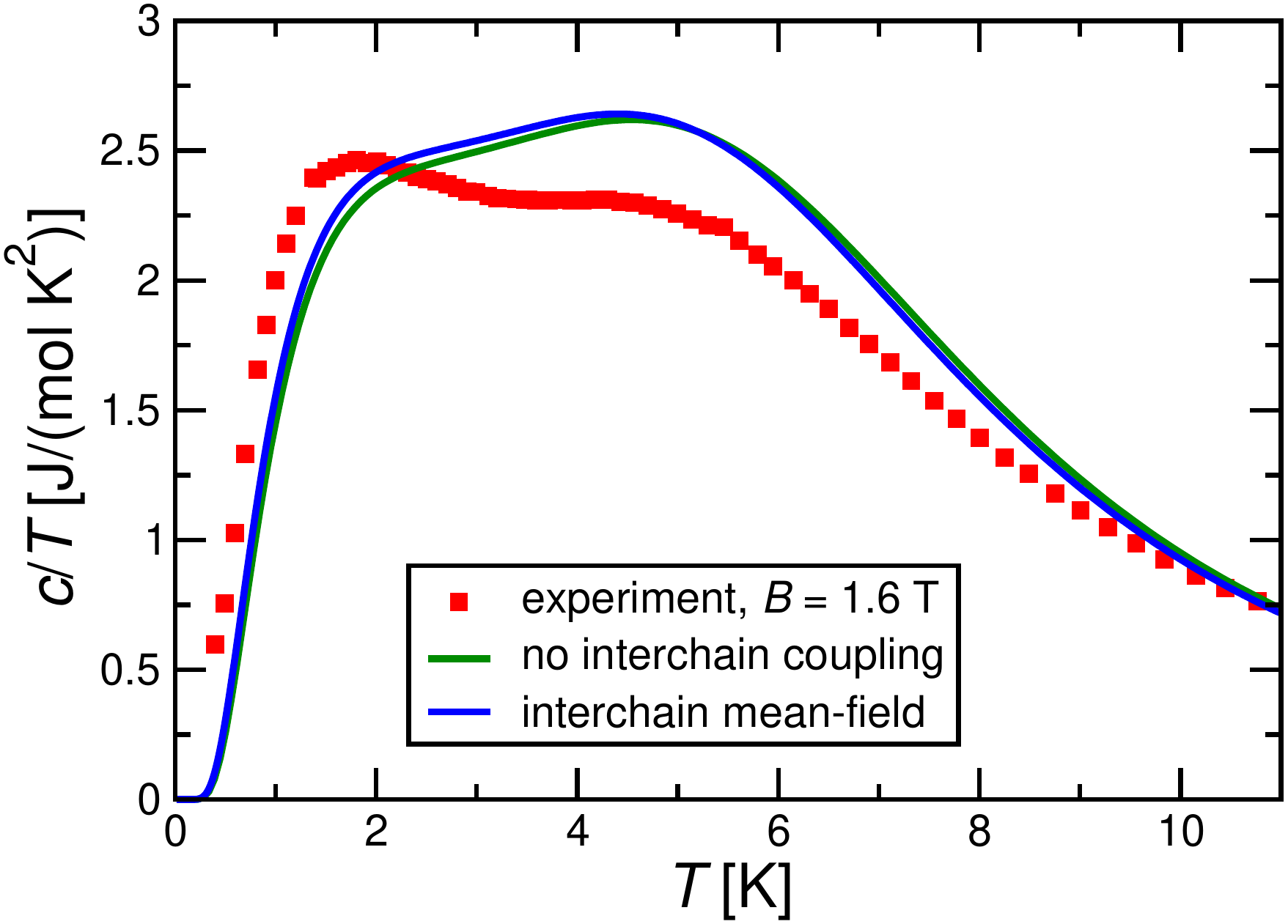}
\end{center}
\vspace*{-3mm}
\caption{\label{fig:CMFTh0.8}
Magnetic specific heat per spin divided by temperature $c/T$ for 
$J=2.8$~K, $D=0.36$~K in a magnetic field $h=0.8\,J$ in comparison with 
the experimental results for $B=1.6$~T. As in Fig.~\ref{fig:CMFTh0.4}, 
theoretical results are obtained by ED with $N=10$ both for decoupled 
chains ($J_\perp=0$) and with a self-consistent mean-field treatment for 
$J_\perp = 0.072$~K.
}
\end{figure}

We are now in a position to perform a comparison with the experimental 
results for the specific heat of Fig.~\ref{fig:fig3}. 
Figures \ref{fig:CMFTh0}--\ref{fig:CMFTh0.8} show the results for $h=0$, 
$0.4\,J$, and $0.8\,J$ (corresponding to the experimental magnetic fields 
$B=0$, $0.8$~T, and $1.6$~T, respectively). For $h=0$ we have used QMC 
with $N=100$ while for $h=0.4\,J$ and $0.8\,J$ we have used ED with 
$N=10$. These systems sizes should be sufficiently large to render 
finite-size effects negligible according to the discussions in 
Sec.~\ref{sec:TheoryDecoupled}. From a technical point of view, we note 
that at $h=0$ and in the paramagnetic phase, $\langle M \rangle=0$ such 
that $\frac{\partial \,\langle M\rangle}{\partial T}=0$ and the 
correction term in Eq.~(\ref{eq:CMF2final}) vanishes, {\it i.e.}, $c_{\rm 
MF} = c_{1D}$, and the blue circles are identical to the green diamonds 
in Fig.~\ref{fig:CMFTh0} for $T>T_N$.

Figures \ref{fig:CMFTh0.4} and \ref{fig:CMFTh0.8} show that the 
interchain coupling leads only to small corrections for a magnetic field 
$h \ge 0.4\,J$; the trend is towards the experimental data, but the shift 
by interchain coupling does not change the situation significantly. 
Nevertheless, the two theory curves and the experimental one in 
Figs.~\ref{fig:CMFTh0.4} and \ref{fig:CMFTh0.8} exhibit double-peak 
structures where the two peaks are located at very similar temperatures 
between theory and experiment.

Figure \ref{fig:CMFTh0} demonstrates that in zero field ($h=0$), 
interchain coupling is not only essential for reproducing the ordering 
transition to good accuracy, as we have seen before, but that thanks to 
the ``Stoner factor'', the correction term in (\ref{eq:CMF2final}) 
dominates the specific heat just below the ordering transition and thus 
gives rise to the characteristic ordering peak. We note, however, that 
the singularity in the denominator of Eq.~(\ref{eq:CMF2final}) is 
deceptive since the numerator (\ref{eq:dMdT}) also vanishes such that $c$ 
has a finite limit for $T \nearrow T_N$. Consequently, our interchain 
mean-field theory remains in the universality class of Landau theory 
\cite{Landau37} with a specific heat exponent $\alpha = 0$.

\section{Magnetocaloric properties}

\label{sec:MCE}

The strong dependence of the specific heat of MnNi(NO$_2$)$_4$(en)$_2$ on 
an applied magnetic field promises a strong magnetocaloric effect and 
potential relevance to low-temperature magnetic refrigeration by 
adiabatic demagnetization, see, e.g., Refs.~\cite{Wolf14,Konieczny22}. 
Therefore, let us have a closer look at its magnetocaloric properties.

\begin{figure}[t!]
\begin{center}
\includegraphics[width=0.69\columnwidth]{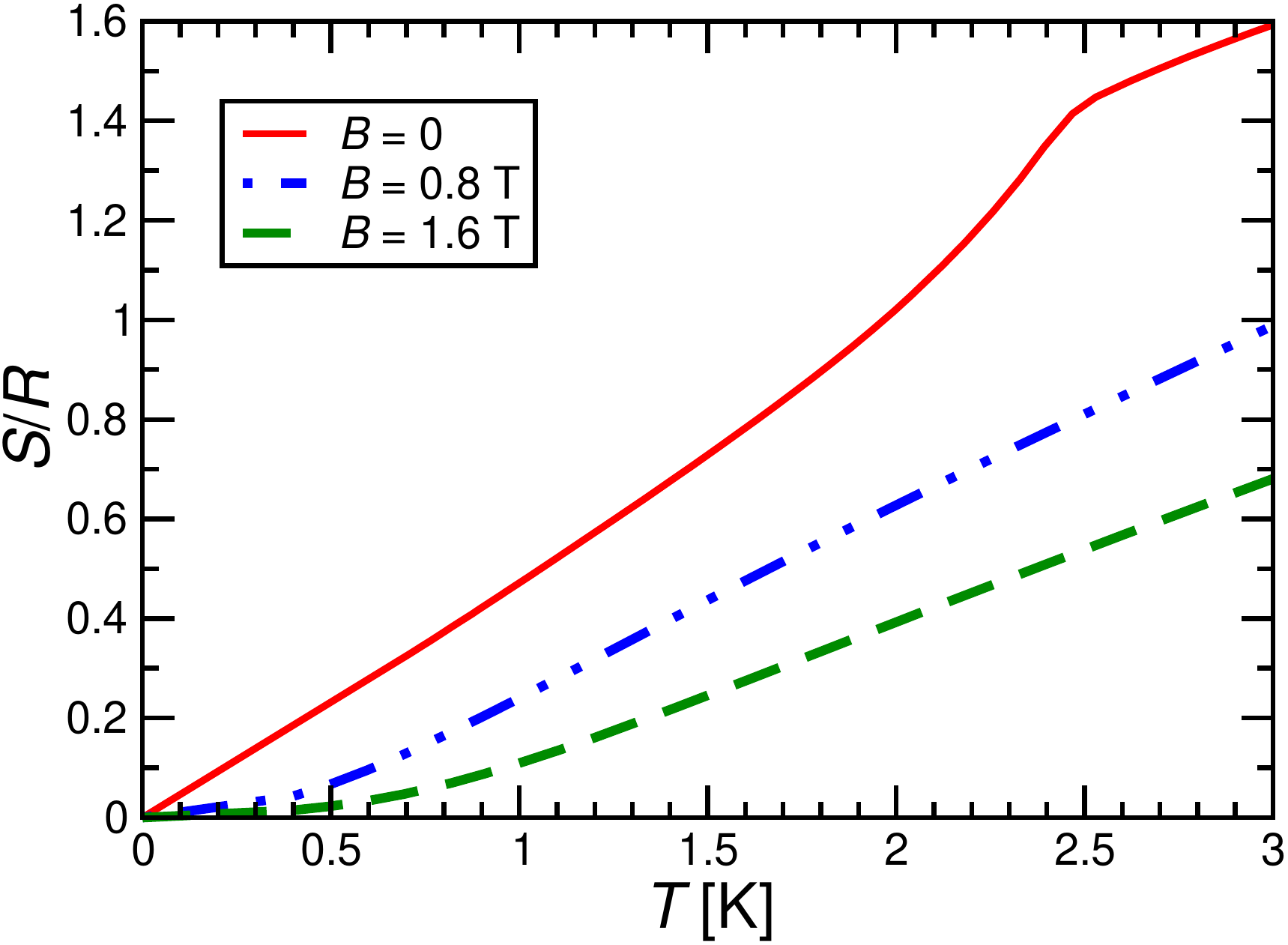}
\end{center}
\vspace*{-3mm}
\caption{\label{fig:Sexp}
Magnetic entropy per mole of MnNi(NO$_2$)$_4$(en)$_2$ in magnetic fields 
of $B=0$, $0.8$, and $1.6$~T, respectively.
}
\end{figure}

Figure \ref{fig:Sexp} shows the experimental magnetic entropy that is 
obtained by integrating the experimental results for the specific heat 
$c_{p,mag}/T$ of Fig.~\ref{fig:fig3}(b) with respect to temperature $T$. 
The $B=0$ curve corresponds to the one shown already in 
Fig.~\ref{fig:fig2}(b). Figure \ref{fig:Sexp} shows that the magnetic 
entropy is significantly reduced by applying a magnetic field of 
$B=1.6$~T, or even $0.8$~T, corresponding to polarization of the spin 
system by the applied magnetic field. Consequently, we expect cooling of 
the spin system during adiabatic demagnetization. Let us consider for 
example an ideal adiabatic process that starts with $T=1.5$~K for 
$B=1.6$~K. We read off from Fig.~\ref{fig:Sexp} that the same entropy is 
found at $B=0$ for $T \approx 0.5$~K, {\it i.e.}, adiabatic 
demagnetization from $B=1.6$~T to $B=0$ would cool from an initial 
temperature $T=1.5$~K to a final temperature of $T \approx 0.5$~K. 
Likewise, an ideal adiabatic process starting with $T=2.5$~K at $B=1.6$~T 
would cool to $T \approx 1.1$~K during a single ideal adiabatic 
demagnetization process. These are relatively large effects in the liquid 
Helium range, which is also remarkable since one is cooling through a 
phase transition into a magnetically ordered state. The main caveat is 
that the processes of the two examples exploit only 8\%\ or 19\%\ of the 
total magnetic entropy $S \approx 2.89\,R$ in the first and second case, 
respectively.

\begin{figure}[t!]
\begin{center}
\includegraphics[width=0.69\columnwidth]{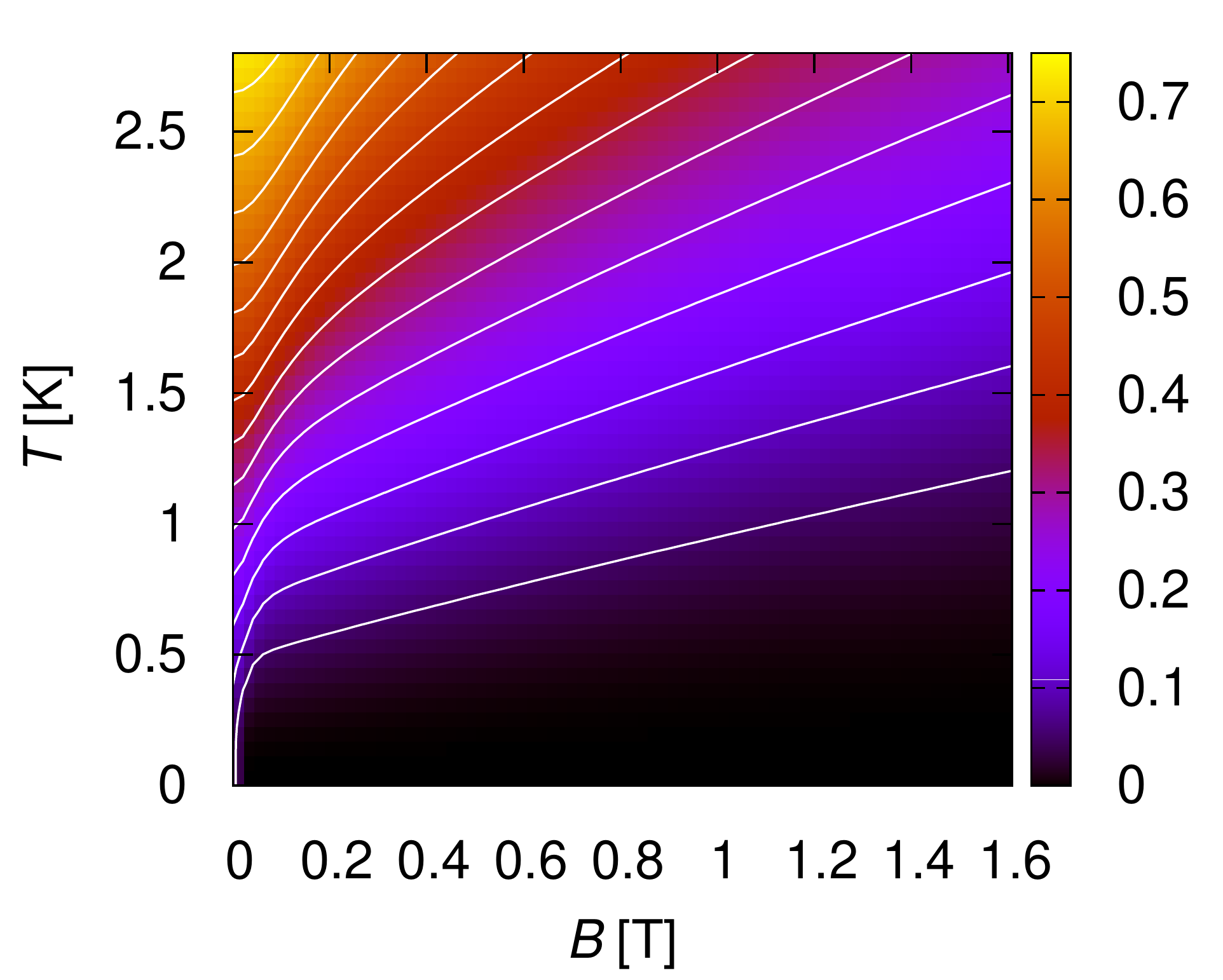}
\end{center}
\vspace*{-3mm}
\caption{\label{fig:Sscan}
Entropy $S$ per spin calculated by ED for an isolated chain with $N=10$ 
spins and $J=2.8$~K, $D=0.36$~K.
}
\end{figure}

Next, let us comment on a numerical description. The entropy is not 
directly accessible in QMC simulations such that we resort to ED even if 
this leads to stronger finite-size effects. Furthermore, for the full $h$ 
and $T$ dependence of the magnetic entropy $S$, we would have to model 
the ordered state in an external magnetic field (grey shaded region in 
the inset of Fig.~\ref{fig:fig3}(b)). However, this is expected to 
correspond to a canted spin configuration and is thus beyond the present 
investigation. We therefore also neglect interchain coupling, {\it i.e.}, 
we focus on a situation corresponding to the one discussed in 
Sec.~\ref{sec:TheoryDecoupled} (see, however, Appendix~\ref{sec:MFT} for 
a discussion of simple single-site mean-field theory). Figure 
\ref{fig:Sscan} shows the corresponding result for the entropy (now 
normalized per spin) of an $N=10$ chain. This density plot of $S(B,T)$ 
permits to immediately read off the magnetocaloric effect. In particular 
the isentropes, corresponding to the white lines in Fig.~\ref{fig:Sscan}, 
directly show the behavior under an adiabatic process. Finite-size 
effects are expected to be small for $B \ge 0.8$~T (corresponding to $h/J 
\ge 0.4$, compare Figs.~\ref{fig:Ch0.4Df} and \ref{fig:Ch0.8Df}), but 
they are known to be relevant throughout the temperature range of 
Fig.~\ref{fig:Sscan} for $B=0$ ($h=0$, compare Fig.~\ref{fig:Ch0Df}).

We also do read off cooling by adiabatic magnetization from 
Fig.~\ref{fig:Sscan} with a size of the effect corresponding to the 
experimental data of Fig.~\ref{fig:Sexp}\footnote{Note that the entropy 
is normalized to mole in Fig.~\ref{fig:Sexp} and per spin in 
Fig.~\ref{fig:Sscan}, amounting to a factor two difference in addition to 
the factor $R$.}. Since we have ferromagnetic chains, the strongest 
cooling occurs in Fig.~\ref{fig:Sscan} upon approaching a zero external 
field. If one adds antiferromagnetic interchain coupling, we expect to 
recover the magnetically ordered phase that is observed experimentally 
for $B < 0.4$~T (compare inset of Fig.~\ref{fig:fig3}(b)) and then 
cooling might actually occur when entering this phase\footnote{Compare 
Fig.~\ref{fig:SscanMFT} and the related discussion in 
Appendix~\ref{sec:MFT} for the behavior in single-site mean-field 
theory.}. Indeed, Fig.~\ref{fig:CMFTh0} demonstrates that interchain 
coupling reshuffles entropy from low temperatures to the ordering 
transition such that the most significant cooling probably occurs around 
it.

\section{Conclusions and Perspectives}

\label{sec:Concl}

We have carried out specific heat measurement in zero and applied field on 
the bimetallic chain compound MnNi(NO$_2$)$_4$(en)$_2$. By determining the 
lattice contribution of the specific heat we have extracted the magnetic 
specific heat $c_{p,mag}$. For the first time, in its temperature 
dependence we verify a long-predicted double-peak like structure. 
Comparison with numerical calculations for the bimetallic $S_1 = 1$, $S_2 
= 5/2$ ferromagnetic spin chain yields a very close resemblance on a 
semi-quantitative level.

Alternating spins are not the only mechanism that may give rise to a 
double-peak structure in the specific heat. For example, also a 
ferromagnetic $S=1$ chain alone can give rise to such structures when 
subjected to a strong single-ion anisotropy $D$ \cite{JIR05}. However, the 
numerical data of Sec.~\ref{sec:TheoryDecoupled} (and further results that 
we do not show here) demonstrate that these two features are already 
present at $D=0$ and can be traced to finite $D$ even if the presence of a 
single-ion anisotropy does affect the behavior of the specific heat at a 
quantitative level. Hence, we conclude that our experimental observation 
of a double-peak like structure in the specific heat directly reflects the 
alternating spins $S_1=1$ and $S_2=5/2$ along the chains. The application 
of an external magnetic field to MnNi(NO$_2$)$_4$(en)$_2$ is essential to 
suppress magnetic order and thus reveal this double-peak feature 
experimentally.

The ordered phase that is observed in MnNi(NO$_2$)$_4$(en)$_2$ for low 
temperatures and small applied magnetic fields is due to an 
antiferromagnetic interchain coupling. Although its absolute value is much 
smaller than the ferromagnetic coupling along the chains, it has a strong 
effect at low temperatures and in the absence of a magnetic field. In 
order to describe this ordered phase, we have developed a mean-field 
treatment of interchain coupling. The combination of QMC simulations for 
isolated chains and such an inter-chain mean-field theory not only yields 
a remarkably accurate value for the ordering transition temperature $T_N$ 
using previously determined parameters \cite{fukushima}, but also yields 
excellent agreement for the full temperature dependence of the magnetic 
specific heat. For fields $h \ge 0.4\,J$, the mean-field corrections are 
small, reflecting the smallness of the interchain coupling constant 
$J_\perp$.

Beyond the very close resemblance on a qualitative level, there are some 
quantitative differences between experiment and theory. For instance, 
while in the calculations the maximum of $c_{p,mag}/T$ is found close to 
$T_{\rm up}$, in the experiments it is observed at $T_{\rm low}$. These 
small difference may be due to the single-ion anisotropy being located 
both on the Ni and Mn sites, and not just the Mn ones, or effects of 
interchain coupling beyond mean-field theory. However, a further 
refinement of the model would require additional information about the 
excitation spectrum such as inelastic neutron scattering.

Another theoretical challenge concerns the theoretical description of the 
ordered state in a magnetic field. For $h=0$ and strong fields along the 
anisotropy axis ($h \ge 0.4$), one may restrict the discussion to 
magnetization along the $z$-axis only. However, for a magnetic field 
applied at an angle to the anisotropy axis, and also for ordered phases 
where the ordered moment cants away from the field/anisotropy axis, it 
will in general be necessary to replace the last term in (\ref{eq:HopMFi}) 
by vectors, {\it i.e.}, by $\left(\vec{h}-\sum_{j\ne i} J_{i,j}\,\langle 
\vec{M}_j\rangle\right)\cdot \sum_{x=1}^{N/2} \left(\vec{S}_x + 
\vec{s}_x\right)$. This generalization can be implemented in single-site 
mean-field theory, but such a strong approximation fails to be 
quantitatively accurate for the present situation (compare 
Appendix~\ref{sec:MFT}). By contrast, implementation of such generic field 
directions in the the interchain mean-field theory of 
Sec.~\ref{sec:ChainMFT} will break conservation of total $S^z$, render the 
computations even more challenging, and thus goes beyond the present 
investigation.

Finally, we have shown that the strong sensitivity of 
MnNi(NO$_2$)$_4$(en)$_2$ to even small applied magnetic fields gives rise 
to a strong magnetocaloric effect, {\it i.e.}, large cooling by adiabatic 
demagnetization from initial fields $B$ on the order of $1$~T. Even if 
the magnetic entropy of MnNi(NO$_2$)$_4$(en)$_2$ may be a bit small for 
practical applications in the temperature range of interest, this 
observation suggests materials with competing strong ferromagnetic and 
weaker antiferromagnetic interactions as promising candidates for 
efficient low-temperature refrigeration.

\authorcontributions{%
R.F.\ provided the samples. Specific heat measurements were performed by M.B.\ and S.S.
W.B.\ and A.H.\ developed the theoretical approach; numerical computations were performed by A.H.
M.T.\ performed the single-site mean-field calculations.
A.H.\ and S.S.\ wrote the manuscript.
All authors have read and agreed to the published version of the manuscript.}

\funding{This works has been supported by the CNRS via the International 
Research Network ``Strongly correlated electron systems as advanced 
magnetocaloric materials’’.}

\acknowledgments{We would like to thank N.\ Fukushima, T.\ Giamarchi, S.\ 
Grossjohann, J.\ Richter, S.\ Wessel, A.U.B.\ Wolter, and M.E.\ 
Zhitomirsky for useful discussions as well as M.\ Meissner for support 
with the heat capacity measurements. Part of the computations have been 
carried out on the ``osaka'' cluster at the Centre de Calcul (CDC) of CY 
Cergy Paris Universit\'e.}

\appendixtitles{yes} 
\appendixstart
\appendix

\section[\appendixname~\thesection]{One-magnon dispersion}

\label{sec:1mag}

\begin{figure}[t!]
\begin{center}
\includegraphics[width=0.69\columnwidth]{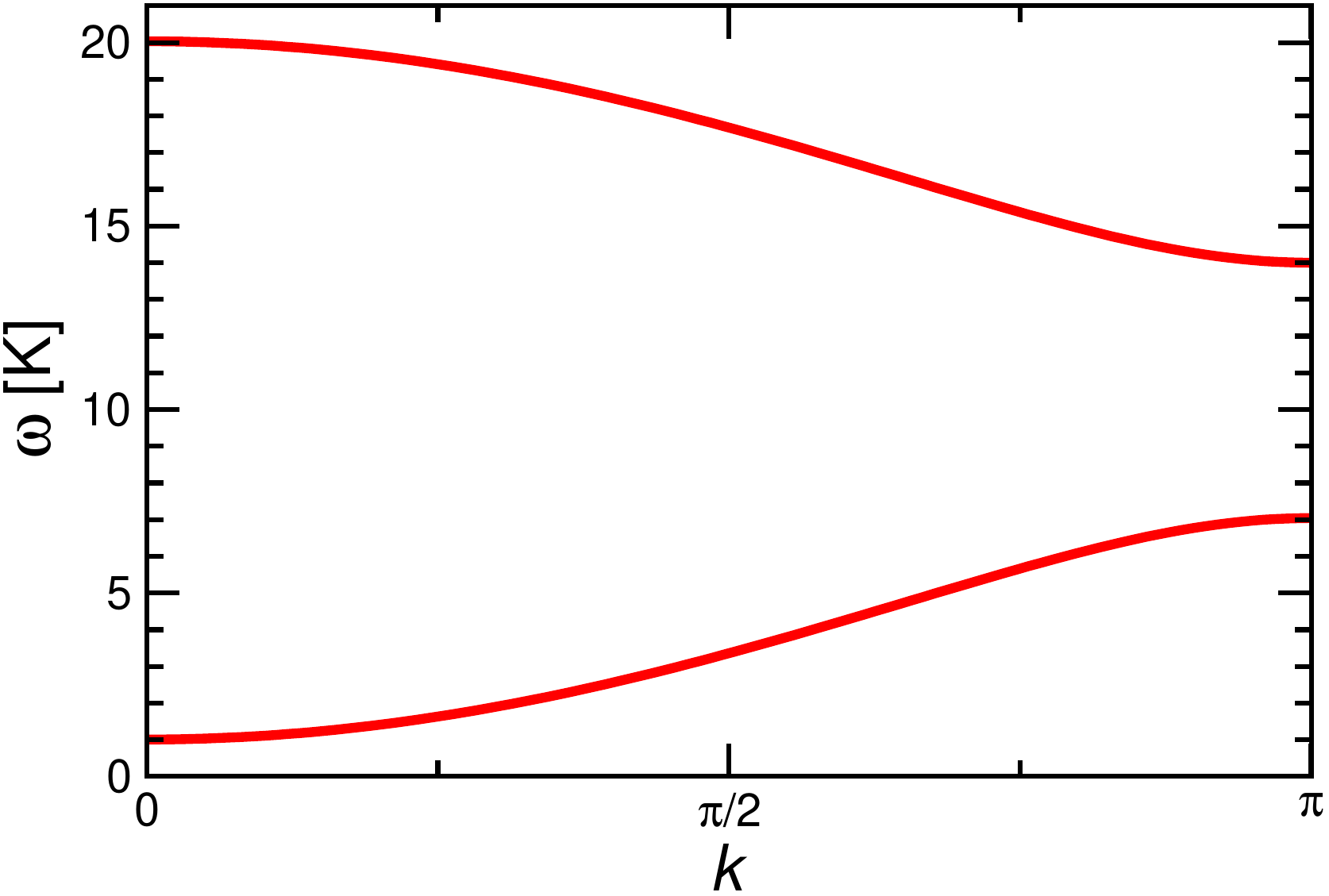}
\end{center}
\vspace*{-3mm}
\caption{\label{fig:omega}
The one-magnon dispersion (\ref{eq:omega}), (\ref{eq:omegaAB}) for
$J=2.8$~K, $D=0.36$~K, and $d=0$.
}
\end{figure}

Let us generalize the computation of the one-magnon dispersion of 
Ref.~\cite{fukushima} to the presence of single-ion anisotropies. To this 
end, we generalize the chain model (\ref{eq:Hop}) to include anisotropy 
terms on both Mn and Ni sites, but drop the magnetic field term:
\begin{equation}
H = -J \sum_{x=1}^{N/2} \left(\vec{S}_x \cdot \vec{s}_x +
\vec{s}_x \cdot \vec{S}_{x+1}\right) 
- D \sum_{x=1}^{N/2} \left(S^z_x\right)^2 
- d \sum_{x=1}^{N/2} \left(s^z_x\right)^2 
 \, .
\label{eq:Hop2D}
\end{equation}
Since the coupling along the chain is ferromagnetic, the ground state is 
also ferromagnetic. A $D$, $d>0$ select the two maximally polarized 
components of the ground state. The one-magnon sector is then obtained by 
flipping a single spin relative to this polarized state. This is a 
single-particle problem that is straightforward to solve by Fourier 
transformation and diagonalization of the $2 \times 2$ matrix resulting 
from the two-site unit cell. This yields two branches of one-magnon 
excitation energies
\begin{eqnarray}
\omega_\pm(k) &\!\!=\!\!&
J\,\left(S_2+S_1\right) + A(S_2) + B(S_1)
\label{eq:omega} \\
&&\pm
\sqrt{J^2\,\left(S_2^2+S_1^2+2\,S_2\,S_1\,\cos(k)\right)
+2\,J\,\Delta(S_1, S_2)\,\left(S_1-S_2\right)
+\Delta(S_1, S_2)^2}
\nonumber
\end{eqnarray}
with
\begin{equation}
A(S_2) = \frac{2\,S_2-1}{2}\, D \, , \qquad
B(S_1) = \frac{2\,S_1-1}{2}\, d \, , \qquad
\Delta(S_1, S_2) = A(S_2)-B(S_1) \, .
\label{eq:omegaAB}
\end{equation}
Figure \ref{fig:omega} shows the two branches of the one-magnon 
dispersion $\omega_\pm(k)$ for the model (\ref{eq:Hop}) and the 
parameters that we have used in the main text. The most important 
qualitative difference to the previous analysis in Ref.~\cite{fukushima} 
is the opening of a gap $\omega_{-}(0) \approx 1$~K due to the single-ion 
anisotropy $D=0.36$~K. Further inspection of (\ref{eq:omega}), 
(\ref{eq:omegaAB}) shows that a reshuffling of the Mn anisotropy to the 
Ni one has a significant effect on the gap $\omega_{-}(0)$. For example, 
the parameters $J=2.8$~K, $D=0$, and $d=1.44$~K would conserve $A(S_2) + 
B(S_1)$ and yield an overall picture that is very similar to 
Fig.~\ref{fig:omega}, but a reduced gap $\omega_{-}(0) \approx 0.4$~K. 
Such a reduction of the gap may indeed be consistent with the 
experimental data in Figs.~\ref{fig:CMFTh0.4} and \ref{fig:CMFTh0.8}, but 
one would need an accurate experimental estimate of the gap for a more 
precise statement.

\section[\appendixname~\thesection]{Details of self-consistency procedure in QMC}

\label{sec:MFTsc}

\begin{figure}[t!]
\begin{center}
\includegraphics[width=0.69\columnwidth]{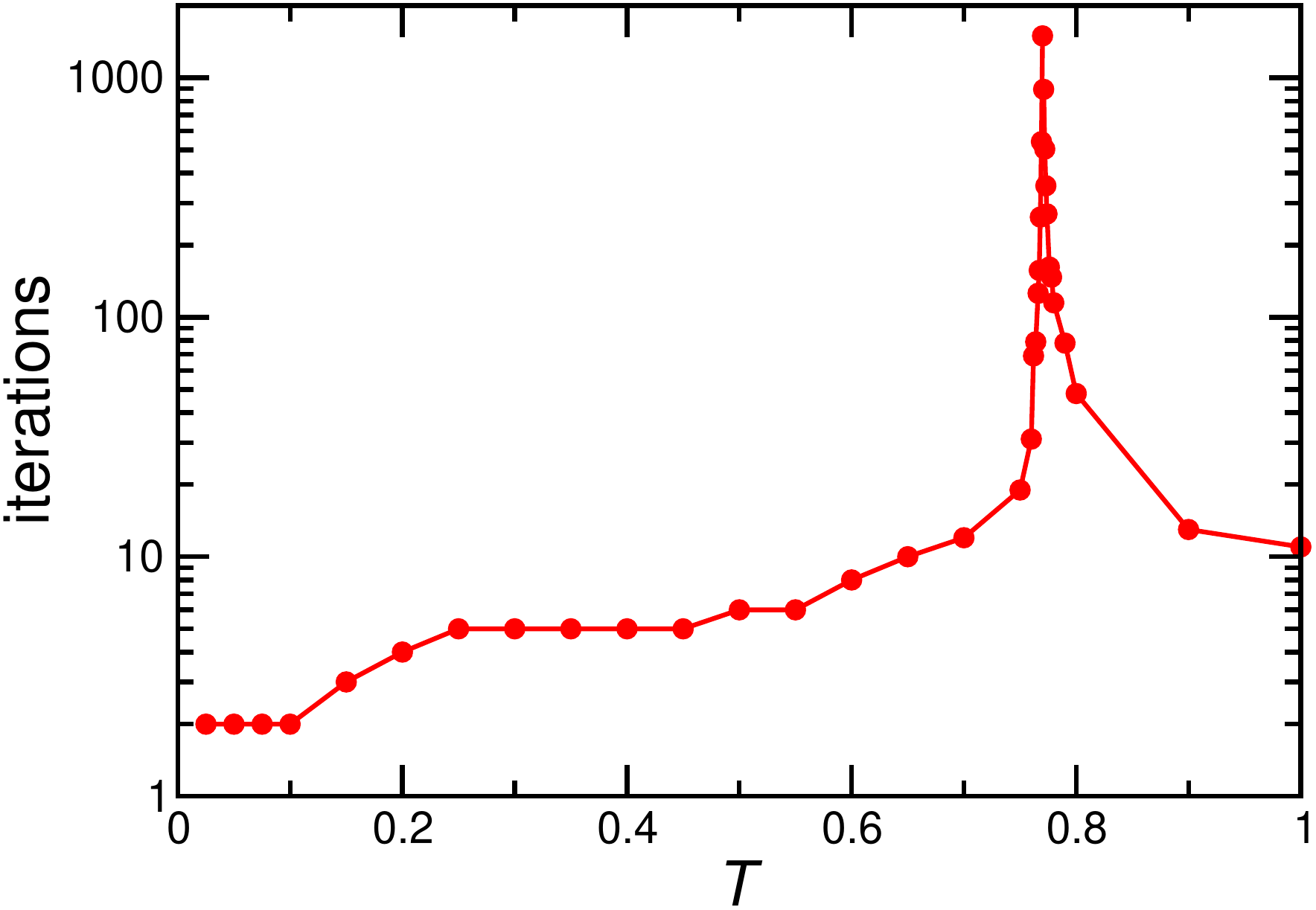}
\end{center}
\vspace*{-3mm}
\caption{\label{fig:MFTit100}
Number of iterations performed for the $N=100$ QMC data
in Fig.~\ref{fig:magMFT} (recall that $J=1$).
Actual data is indicated by filled circles; lines are guides to the eye.
}
\end{figure}

Figure \ref{fig:MFTit100} shows the number of iterations performed in 
order to reach self-consistency for the $N=100$ QMC data presented in 
Fig.~\ref{fig:magMFT}. The precise number of iterations depends on 
details such as the desired level of accuracy (we aimed at reducing the 
error on $\langle M \rangle$ to below $10^{-3}$) and the exact way the 
iterations are run. One can nevertheless draw some qualitative 
conclusions: Sufficiently far away from the N\'eel temperature, 
self-consistency is obtained after a few iterations, but the number of 
required iterations explodes upon approaching the phase transition, a 
phenomenon that may be interpreted as a form of ``critical slowing 
down''. Bearing in mind that it may take a few days to obtain 
sufficiently small statistical error bars within a single iteration and 
that several hundred to more than 1000 iterations have been performed, it 
is also evident that the computations have been running over an extended 
period of time. The procedure could be sped up by a more sophisticated 
root-finding algorithm than simple iteration, but we stayed with the 
latter for the present investigation.

\section[\appendixname~\thesection]{Single-site mean-field approximation}

\label{sec:MFT}

In order to theoretically explore the ordered phase in a small external 
magnetic fields where the spins are expected to be canted, we summarize 
here a complete single-site mean-field decoupling of the Hamiltonian 
(\ref{eq:Hop}) supplemented by the interchain coupling (\ref{eq:Hop1}); 
for further details we refer to chapter 5 of Ref.~\cite{TiwariPhD}.

Since the days of Pierre Weiss \cite{weiss07}, the mean-field 
approximation has become a textbook method in the theory of magnetism 
(see, e.g., Refs.~\cite{Blundell01,PiresTeixeira21} and references 
therein) such that we will comment only briefly on it. The essential step 
is to replace the terms in Eq.~(\ref{eq:Hop}) as follows:
\begin{equation}
\vec{S}_x \cdot \vec{s}_y
\ \rightarrow \
\vec{S}_x \cdot \langle \vec{s}_y \rangle
+ \langle \vec{S}_x \rangle \cdot \vec{s}_y
- \langle \vec{S}_x \rangle \cdot \langle \vec{s}_y \rangle \, .
\label{eq:MFTdec}
\end{equation}
In combination with the mean-field decoupling of the interchain coupling 
(\ref{eq:HopMFi}) this leads to a set of single-spin problems with 
individual coupling to the external magnetic field, possibly single-ion 
anisotropy, and coupling to their neighbors taken into account effectively 
via an additional mean field. However, the expectation values $\langle 
\vec{S}_x \rangle$ and $\langle \vec{s}_y \rangle$ need to be determined 
self-consistently for this set of coupled problems. We solve this 
self-consistency condition by iteration, {\it i.e.}, we assume a 
configuration of the $\langle \vec{S}_x \rangle$ and $\langle \vec{s}_y 
\rangle$, solve the single-ion problems numerically, recompute the 
expectation values, and iterate until convergence. In principle, the 
procedure can be implemented for a lattice of coupled mean-field problems 
(see, e.g., Refs.~\cite{Melchy09,GZZ16}). However, we make some further 
plausible assumption in order to reduce the numerical effort. Firstly, in 
view of the ferromagnetic coupling along the chain, we assume the pattern 
to be translationally invariant although we do need two mean fields due to 
the alternating spins. Secondly, in view of the antiferromagnetic 
interchain coupling, we allow for two inequivalent chains. This leads to a 
set of four mean-field coupled single-ion problems. Finally, we assume the 
spin configuration to lie in a plane that includes the external magnetic 
field (and thus also the single-ion anisotropy that we assume to be 
parallel to the magnetic field).

It turns out that there is no finite-field phase in the parameter regime 
studied in the main text. This may be attributed to the antiferromagnetic 
interchain coupling $J_\perp$ just partially cancelling the ferromagnetic 
chain coupling $J$ when all couplings are treated at the mean-field level, 
thus leading to an effectively ferromagnetic system with just a reduced 
effective coupling constant. We therefore use modified parameters $J=1$, 
$J_\perp=0.5$, $D=0.1$ in this appendix and focus on the qualitative 
behavior.

\subsection{Phase diagram}

\begin{figure}[t!]
\begin{center}
\includegraphics[width=0.49\columnwidth]{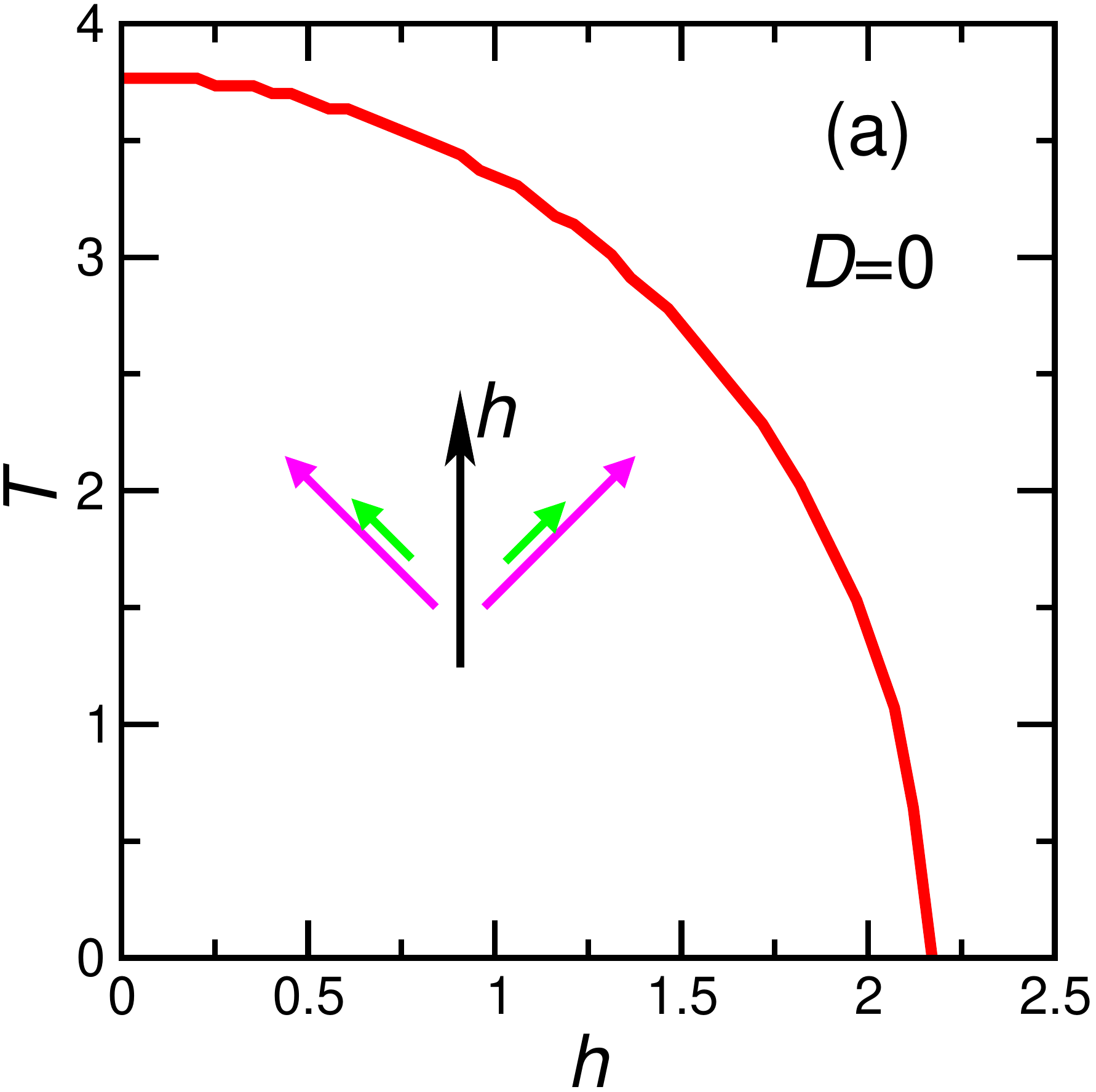}\hfill
\includegraphics[width=0.49\columnwidth]{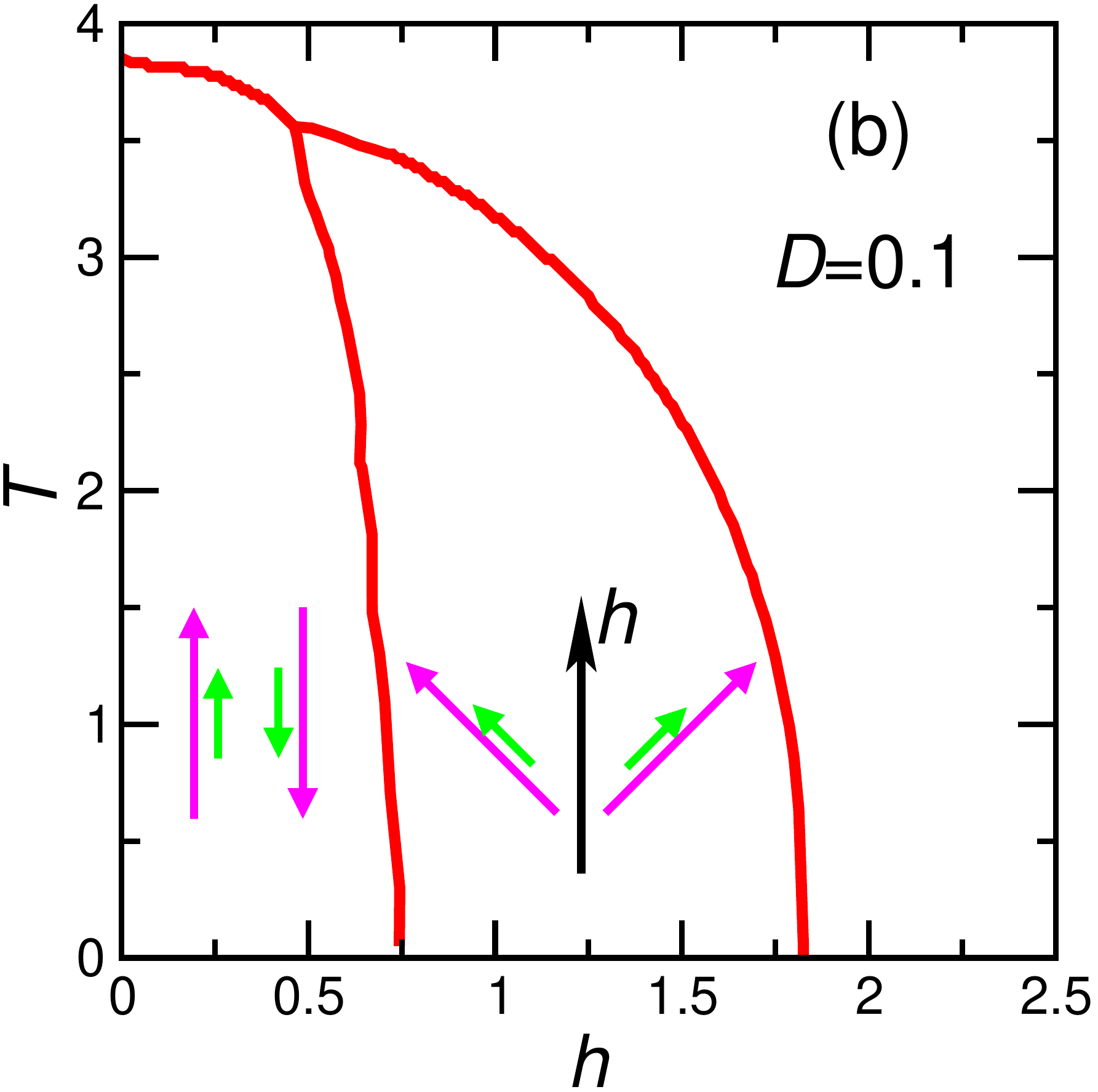}
\end{center}
\vspace*{-3mm}
\caption{\label{fig:MFTphasediag}
Single-site mean-field phase diagrams for $J=1$, $J_\perp=0.5$ and $D=0$ 
(a) and $D=0.1$ (b). The schematics indicate the spin configurations 
relative to the applied magnetic field in the ordered phases. A pair of 
parallel arrows indicates Mn (magenta) and Ni (green) spins in one chain, 
the second pair indicates the neighboring chain. The phase at high 
temperature or large magnetic fields is paramagnetic.
}
\end{figure}

Figure \ref{fig:MFTphasediag} shows the mean-field phase diagram with and 
without the single-ion anisotropy. The phase diagrams were obtained from 
an analysis of the spin configurations \cite{TiwariPhD}. In the case of 
$D=0$ (Fig.~\ref{fig:MFTphasediag}(a)) we find an ordered 
antiferromagnetic phase. At zero field $h=0$, the direction of the 
ordering vector is arbitrary. Application of a small field orients the 
spins orthogonal to the field direction. Upon increasing magnetic field, 
spins are increasingly tilted towards the field direction. The overall 
behavior is very similar to the well-known Heisenberg antiferromagnet on 
a bipartite lattice (see, e.g., Ref.~\cite{ZN98} for the case of the 
square lattice), the main difference being that all Mn and Ni spins in 
one chain adopt the role of those of one sublattice. Note that in the 
case of $D=0$, the Mermin-Wagner theorem \cite{MW66} would forbid a 
finite-temperate transition in the case of one or even two dimensions. 
The phase diagram of Fig.~\ref{fig:MFTphasediag}(a) should thus be 
thought of to represent the case of chains coupled in three dimensions.

Figure \ref{fig:MFTphasediag}(b) presents the phase diagram for a 
single-ion anisotropy $D>0$. The main difference with the $D=0$ case is 
the appearance of an additional phase at small magnetic fields. Indeed, 
the single-ion anisotropy pins the spins along the anisotropy axis. For a 
small magnetic field (that we choose here to be parallel to the 
anisotropy axis), the spins remain pinned along this axis and a finite 
critical field is needed to enter the ``spin-flop'' phase where the spins 
cant towards the magnetic field and that we already observed for the case 
$D=0$ (Fig.~\ref{fig:MFTphasediag}(a)). The structure of the phase 
diagram Fig.~\ref{fig:MFTphasediag}(b) is again reminiscent of the 
well-known phase diagram of an anisotropic antiferromagnet on a bipartite 
lattice (see, e.g., 
Refs.~\cite{Fisher75,LB78,HassaniMaster,Blundell01,Selke09}). The main 
difference is again that the Ni ($S_1=1$) and Mn ($S_2 = 5/2$) spins of 
one chain pair up to correspond to one sublattice. Actually, when the 
spins tilt with respect to the magnetic field, Ni and Mn ones are not 
expected to be exactly parallel to each other, in particular if the Mn 
one is subject to a single-ion anisotropy while the Ni one is not. 
However, it turns out that the angle between a pair does not exceed a few 
degrees \cite{TiwariPhD}. Accordingly, the sketches of the spin 
configurations in Fig.~\ref{fig:MFTphasediag} are schematic in the sense 
that spin pairs are almost but not necessarily exactly parallel. For 
$D>0$, the Mermin-Wagner theorem \cite{MW66} allows finite-temperature 
ordering starting in two dimensions. Nevertheless, for small values of 
$D$ and weakly coupled ferromagnetic chains, {\it i.e.}, the situation 
relevant to MnNi(NO$_2$)$_4$(en)$_2$, one is still close to a situation 
where ordering would be forbidden such that the present mean-field theory 
is likely to overestimate the transition temperature. Indeed, at $h=0$, 
the estimate inferred from Fig.~\ref{fig:chiTN} is $T_N < J$ rather than 
$T_N \approx 4\,J$, as observed in Fig.~\ref{fig:MFTphasediag}.

Let us briefly comment on a comparison to the experimental phase diagram. 
The inset of Fig.~\ref{fig:fig3} just shows the transition into an 
ordered phase. However, the experimental data for the magnetization and 
Bragg intensity of elastic neutron scattering show two features at 
$T=1.8$~K as a function of applied field $B$ \cite{feyerherm}. We believe 
that these two experimental features correspond to the two transitions in 
the mean-field phase diagram Fig.~\ref{fig:MFTphasediag}(b).

\subsection{Entropy and magnetocaloric effect}

\begin{figure}[t!]
\leftline{\hskip0.02\columnwidth(a)\hskip0.43\columnwidth(b)}
\vspace*{-2mm}
\begin{center}
\includegraphics[height=0.43\columnwidth]{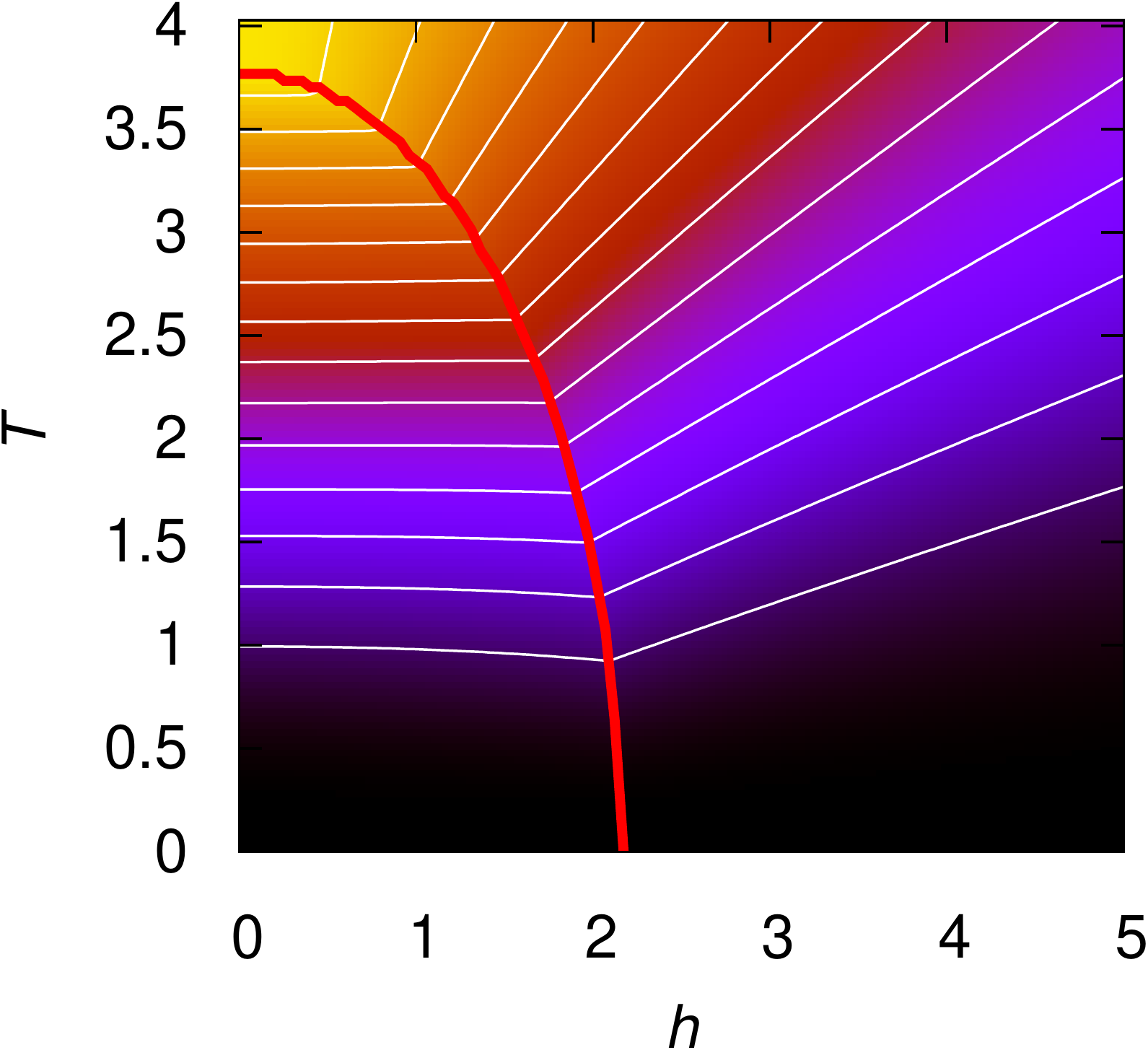}\hfill
\includegraphics[height=0.43\columnwidth]{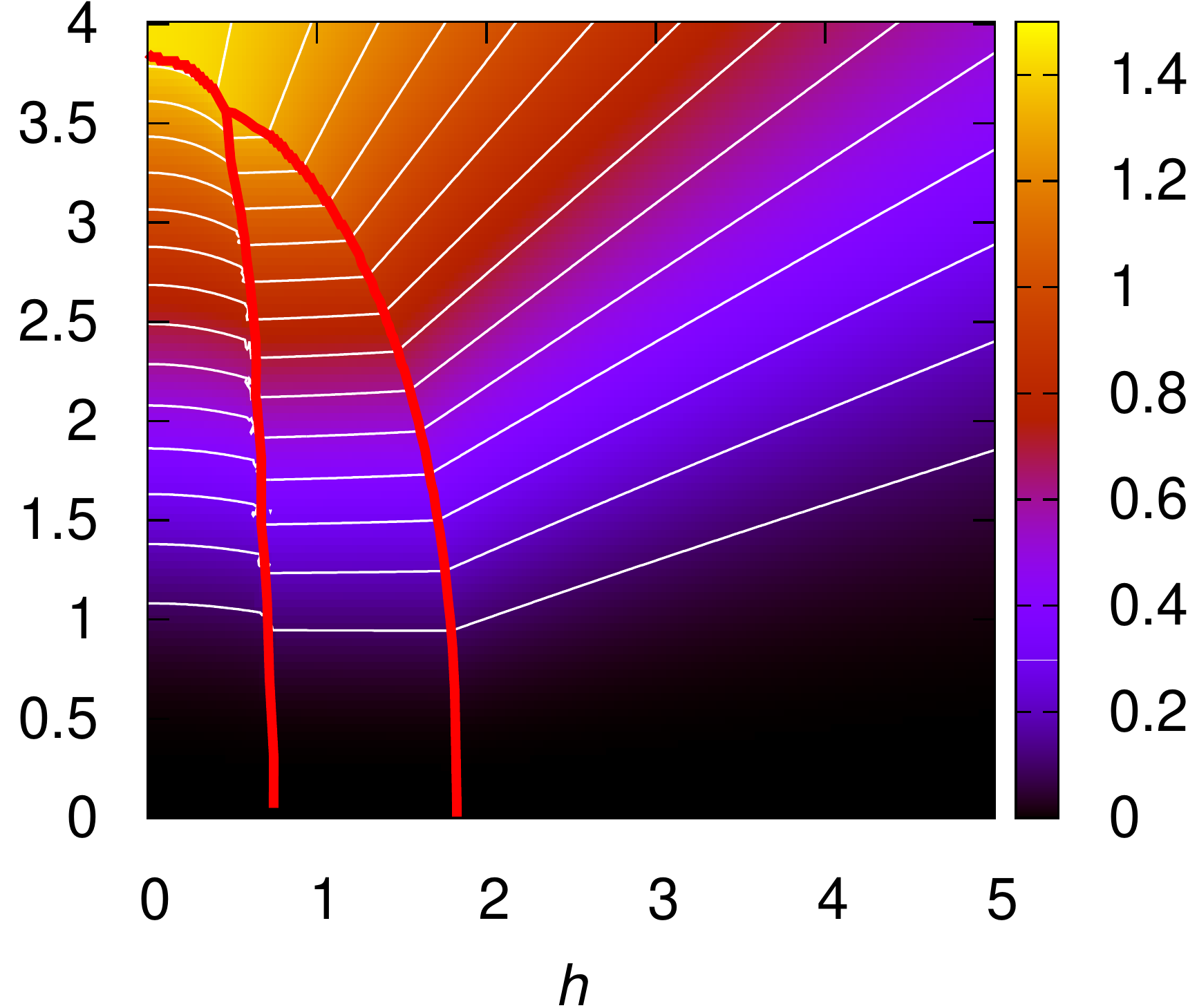}
\end{center}
\vspace*{-1mm}
\caption{\label{fig:SscanMFT}
Single-site mean-field results for the entropy per spin at $J=1$, 
$J_\perp=0.5$ and $D=0$ (a) and $D=0.1$ (b). Red lines are the phase 
boundaries from Fig.~\ref{fig:MFTphasediag}.
}
\end{figure}

Now we turn to the magnetocaloric properties. Mean-field theory has been 
used before for this purpose (see, e.g., 
Refs.~\cite{TishinSpichkin,Heydarinasab2020,Palacios2022}). Indeed, once 
the self-consistent mean-field solution is known, both the free energy $F$ 
and the internal energy $U$ are straightforward to compute and from these 
one obtains the entropy via
\begin{equation}
S  = \frac{U-F}{T} \, .
\label{eq:Thermo}
\end{equation}
Figure \ref{fig:SscanMFT} presents results of the entropy for our 
mixed-spin system with $D=0$ (a) and $D=0.1$ (b). The representation is 
analogous to Fig.~\ref{fig:Sscan} with entropy being normalized per spin 
and white lines denoting isentropes. Furthermore, we superimpose the phase 
transitions from Fig.~\ref{fig:MFTphasediag} as red lines in 
Fig.~\ref{fig:SscanMFT}. One observes that the phase boundaries correspond 
to kinks in the isentropes. In fact, the isentropes of 
Fig.~\ref{fig:SscanMFT}(b) are discontinuous across the transition 
separating the collinear antiferromagnetic phase at low magnetic fields 
and the spin-flop phase at higher fields, reflecting the first-order 
nature of this transition.

The main qualitative finding is that we observe cooling by adiabatic 
demagnetization, as expected. However, the effect is mainly restricted to 
the paramagnetic phase. Upon entering the ordered (spin-flop) phase, the 
isentropes become very flat, {\it i.e.}, temperature $T$ varies very 
little when $h$ is varied in this ordered phase. There is even a small 
heating effect when $h$ is lowered through the transition between the 
spin-flop and collinear ordered phase that appears for $D=0.1$ 
(Fig.~\ref{fig:SscanMFT}(b)). A quantitative comparison with 
MnNi(NO$_2$)$_4$(en)$_2$ is unfortunately precluded, e.g., by mean-field 
theory overestimating ordering tendencies, as discussed before in the 
context of Fig.~\ref{fig:MFTphasediag}. Let us note a final peculiarity of 
mean-field theory, namely that the total entropy of the system is 
essentially recovered for $T>T_N$ at $h=0$ and thus visible in 
Fig.~\ref{fig:SscanMFT}\footnote{%
To be precise, symmetry arguments imply that the mean-field correction to 
the specific heat $c$ vanishes at $h=0$ for $T>T_N$, as already stated in 
Sec.~\ref{sec:SecCompExp}. As a consequence, the total entropy is 
recovered exactly for $D=0$. For $D\ne 0$, the single-ion splitting leads 
to some fluctuations surviving for $T>T_N$. However, for $D=0.1\,J$, the 
effect is so small that less than 1 permille of the total entropy is 
missing at the highest temperature shown in Fig.~\ref{fig:SscanMFT}(b).
}
while it would be recovered only in the limit $T \to \infty$ for the case 
of the ED result of Fig.~\ref{fig:Sscan}. We recall furthermore that also 
the experiment recovers barely half of the total entropy at $T_N$ for 
$B=0$, compare Fig.~\ref{fig:fig2}(b)) and the related discussion in 
Sec.~\ref{sec:SpecHeat}.

Let us conclude this discussion by mentioning that the magnetic 
susceptibility $\chi$ and specific heat $c$ can in principle also be 
investigated within the single-site mean-field approximation. However, if 
one wants to avoid numerical derivatives, the presence of two inequivalent 
sites in each chain requires matrix generalizations of 
Eqs.~(\ref{eq:chiRPA}) and (\ref{eq:CMF2final}). We refer to 
Ref.~\cite{TiwariPhD} for further details on these aspects.

\begin{adjustwidth}{-\extralength}{0cm}

\reftitle{References}


\bibliography{MnNi}

\begin{thebibliography}{999}

\bibitem[Mikeska and Kolezhuk(2004)]{Mikeska2004}
Mikeska, H.J.; Kolezhuk, A.K., One-dimensional magnetism.
\newblock In {\em Quantum Magnetism}; Schollw{\"o}ck, U.; Richter, J.; Farnell,
  D.J.J.; Bishop, R.F., Eds.; Springer: Berlin, Heidelberg,  2004; pp. 1--83.
\newblock {\url{https://doi.org/10.1007/BFb0119591}}.

\bibitem[Dender \em{et~al.}(1997)Dender, Hammar, Reich, Broholm, and
  Aeppli]{Dender97}
Dender, D.C.; Hammar, P.R.; Reich, D.H.; Broholm, C.; Aeppli, G.
\newblock Direct Observation of Field-Induced Incommensurate Fluctuations in a
  One-Dimensional {$S=1/2$} Antiferromagnet.
\newblock {\em Phys. Rev. Lett.} {\bf 1997}, {\em 79},~1750--1753.
\newblock {\url{https://doi.org/10.1103/PhysRevLett.79.1750}}.

\bibitem[Oshikawa and Affleck(1997)]{Oshikawa97}
Oshikawa, M.; Affleck, I.
\newblock Field-Induced Gap in {$S=1/2$} Antiferromagnetic Chains.
\newblock {\em Phys. Rev. Lett.} {\bf 1997}, {\em 79},~2883--2886.
\newblock {\url{https://doi.org/10.1103/PhysRevLett.79.2883}}.

\bibitem[Essler and Tsvelik(1998)]{Essler98}
Essler, F.H.L.; Tsvelik, A.M.
\newblock Dynamical magnetic susceptibilities in copper benzoate.
\newblock {\em Phys. Rev. B} {\bf 1998}, {\em 57},~10592--10597.
\newblock {\url{https://doi.org/10.1103/PhysRevB.57.10592}}.

\bibitem[Zvyagin \em{et~al.}(2004)Zvyagin, Kolezhuk, Krzystek, and
  Feyerherm]{zvyagin}
Zvyagin, S.A.; Kolezhuk, A.K.; Krzystek, J.; Feyerherm, R.
\newblock Excitation Hierarchy of the Quantum Sine-{G}ordon Spin Chain in a
  Strong Magnetic Field.
\newblock {\em Phys. Rev. Lett.} {\bf 2004}, {\em 93},~027201.
\newblock {\url{https://doi.org/10.1103/PhysRevLett.93.027201}}.

\bibitem[Tiegel \em{et~al.}(2016)Tiegel, Honecker, Pruschke, Ponomaryov,
  Zvyagin, Feyerherm, and Manmana]{Tiegel16}
Tiegel, A.C.; Honecker, A.; Pruschke, T.; Ponomaryov, A.; Zvyagin, S.A.;
  Feyerherm, R.; Manmana, S.R.
\newblock Dynamical properties of the sine-{G}ordon quantum spin magnet {Cu-PM}
  at zero and finite temperature.
\newblock {\em Phys. Rev. B} {\bf 2016}, {\em 93},~104411.
\newblock {\url{https://doi.org/10.1103/PhysRevB.93.104411}}.

\bibitem[Hagiwara \em{et~al.}(2005)Hagiwara, Regnault, Zheludev, Stunault,
  Metoki, Suzuki, Suga, Kakurai, Koike, Vorderwisch, and Chung]{hagiwara}
Hagiwara, M.; Regnault, L.P.; Zheludev, A.; Stunault, A.; Metoki, N.; Suzuki,
  T.; Suga, S.; Kakurai, K.; Koike, Y.; Vorderwisch, P.;  et~al.
\newblock Spin Excitations in an Anisotropic Bond-Alternating Quantum {$S=1$}
  Chain in a Magnetic Field: Contrast to {H}aldane Spin Chains.
\newblock {\em Phys. Rev. Lett.} {\bf 2005}, {\em 94},~177202.
\newblock {\url{https://doi.org/10.1103/PhysRevLett.94.177202}}.

\bibitem[Wolfram and Ellialtioglu(1980)]{wolfram}
Wolfram, T.; Ellialtioglu, S.
\newblock Neutron Scattering by Magnons of an Antiferromagnet with Modulated
  Spin Amplitudes.
\newblock {\em Phys. Rev. Lett.} {\bf 1980}, {\em 44},~1295--1298.
\newblock {\url{https://doi.org/10.1103/PhysRevLett.44.1295}}.

\bibitem[Bartolom\'e \em{et~al.}(1996)Bartolom\'e, Bartolom\'e, Benelli,
  Caneschi, Gatteschi, Paulsen, Pini, Rettori, Sessoli, and
  Volokitin]{bartolome}
Bartolom\'e, F.; Bartolom\'e, J.; Benelli, C.; Caneschi, A.; Gatteschi, D.;
  Paulsen, C.; Pini, M.G.; Rettori, A.; Sessoli, R.; Volokitin, Y.
\newblock Effect of Chiral Domain Walls on the Specific Heat of
  {Gd}(hfac)$_3${NIT$R$} ({$R$=Ethyl, Isopropyl, Methyl, Phenyl}) Molecular
  Magnetic Chains.
\newblock {\em Phys. Rev. Lett.} {\bf 1996}, {\em 77},~382--385.
\newblock {\url{https://doi.org/10.1103/PhysRevLett.77.382}}.

\bibitem[Drillon \em{et~al.}(1989)Drillon, Coronado, Georges, Gianduzzo, and
  Curely]{drillon}
Drillon, M.; Coronado, E.; Georges, R.; Gianduzzo, J.C.; Curely, J.
\newblock Ferrimagnetic {H}eisenberg chains [(1/2-{$S$}] ({$S=1$} to (5/2):
  Thermal and magnetic properties.
\newblock {\em Phys. Rev. B} {\bf 1989}, {\em 40},~10992--10998.
\newblock {\url{https://doi.org/10.1103/PhysRevB.40.10992}}.

\bibitem[Pati \em{et~al.}(1997{\natexlab{a}})Pati, Ramasesha, and
  Sen]{Pati1997a}
Pati, S.K.; Ramasesha, S.; Sen, D.
\newblock Low-lying excited states and low-temperature properties of an
  alternating spin-1--spin-1/2 chain: A density-matrix renormalization-group
  study.
\newblock {\em Phys. Rev. B} {\bf 1997}, {\em 55},~8894--8904.
\newblock {\url{https://doi.org/10.1103/PhysRevB.55.8894}}.

\bibitem[Pati \em{et~al.}(1997{\natexlab{b}})Pati, Ramasesha, and
  Sen]{Pati1997b}
Pati, S.K.; Ramasesha, S.; Sen, D.
\newblock A density matrix renormalization group study of low-energy
  excitations and low-temperature properties of alternating spin systems.
\newblock {\em J. Phys.: Condens. Matter} {\bf 1997}, {\em 9},~8707--8726.
\newblock {\url{https://doi.org/10.1088/0953-8984/9/41/016}}.

\bibitem[Ivanov(2000)]{Ivanov2000}
Ivanov, N.B.
\newblock Magnon dispersions in quantum {H}eisenberg ferrimagnetic chains at
  zero temperature.
\newblock {\em Phys. Rev. B} {\bf 2000}, {\em 62},~3271--3278.
\newblock {\url{https://doi.org/10.1103/PhysRevB.62.3271}}.

\bibitem[Kolezhuk \em{et~al.}(1997)Kolezhuk, Mikeska, and
  Yamamoto]{Kolezhuk1997}
Kolezhuk, A.K.; Mikeska, H.J.; Yamamoto, S.
\newblock Matrix-product-states approach to {H}eisenberg ferrimagnetic spin
  chains.
\newblock {\em Phys. Rev. B} {\bf 1997}, {\em 55},~R3336--R3339.
\newblock {\url{https://doi.org/10.1103/PhysRevB.55.R3336}}.

\bibitem[Yamamoto and Fukui(1998)]{Yamamoto1998a}
Yamamoto, S.; Fukui, T.
\newblock Thermodynamic properties of {H}eisenberg ferrimagnetic spin chains:
  Ferromagnetic-antiferromagnetic crossover.
\newblock {\em Phys. Rev. B} {\bf 1998}, {\em 57},~R14008--R14011.
\newblock {\url{https://doi.org/10.1103/PhysRevB.57.R14008}}.

\bibitem[Yamamoto \em{et~al.}(1998)Yamamoto, Fukui, Maisinger, and
  Schollw\"ock]{Yamamoto1998b}
Yamamoto, S.; Fukui, T.; Maisinger, K.; Schollw\"ock, U.
\newblock Combination of ferromagnetic and antiferromagnetic features in
  {H}eisenberg ferrimagnets.
\newblock {\em J. Phys.: Condens. Matter} {\bf 1998}, {\em 10},~11033--11048.
\newblock {\url{https://doi.org/10.1088/0953-8984/10/48/023}}.

\bibitem[Nakanishi and Yamamoto(2002)]{Nakanishi2002}
Nakanishi, T.; Yamamoto, S.
\newblock Intrinsic double-peak structure of the specific heat in
  low-dimensional quantum ferrimagnets.
\newblock {\em Phys. Rev. B} {\bf 2002}, {\em 65},~214418.
\newblock {\url{https://doi.org/10.1103/PhysRevB.65.214418}}.

\bibitem[Yamamoto and Hori(2005)]{Yamamoto2005}
Yamamoto, S.; Hori, H.
\newblock Low-temperature thermodynamics of one-dimensional alternating-spin
  {H}eisenberg ferromagnets.
\newblock {\em Phys. Rev. B} {\bf 2005}, {\em 72},~054423.
\newblock {\url{https://doi.org/10.1103/PhysRevB.72.054423}}.

\bibitem[Kahn(1993)]{kahn}
Kahn, O.
\newblock {\em Molecular magnetism}; Wiley-VCH: New York,  1993.

\bibitem[Caneschi \em{et~al.}(1989)Caneschi, Gatteschi, Renard, Rey, and
  Sessoli]{Caneschi89}
Caneschi, A.; Gatteschi, D.; Renard, J.P.; Rey, P.; Sessoli, R.
\newblock Magnetic coupling in zero- and one-dimensional magnetic systems
  formed by nickel({II}) and nitronyl nitroxides. Magnetic phase transition of
  a ferrimagnetic chain.
\newblock {\em Inorg. Chem.} {\bf 1989}, {\em 28},~2940--2944.
\newblock {\url{https://doi.org/10.1021/ic00314a013}}.

\bibitem[Zhou \em{et~al.}(1994)Zhou, Makivic, Zuo, Zane, Miller, and
  Epstein]{Zhou94}
Zhou, P.; Makivic, M.; Zuo, F.; Zane, S.; Miller, J.S.; Epstein, A.J.
\newblock Ferromagnetic behavior and magnetic excitations in a molecular-based
  alternating-spin chain: Decamethylchromocenium tetracyanoethanide.
\newblock {\em Phys. Rev. B} {\bf 1994}, {\em 49},~4364--4367.
\newblock {\url{https://doi.org/10.1103/PhysRevB.49.4364}}.

\bibitem[Nishizawa \em{et~al.}(2000)Nishizawa, Shiomi, Sato, Takui, Itoh, Sawa,
  Kato, Sakurai, Izuoka, and Sugawara]{Nishizawa2000}
Nishizawa, M.; Shiomi, D.; Sato, K.; Takui, T.; Itoh, K.; Sawa, H.; Kato, R.;
  Sakurai, H.; Izuoka, A.; Sugawara, T.
\newblock Evidence for the Breakdown of Simple Classical Pictures of Organic
  Molecule-Based Ferrimagnetics: Low-Temperature Crystal Structure and
  Single-Crystal {ESR} Studies of an Organic Heterospin System.
\newblock {\em J. Phys. Chem. B} {\bf 2000}, {\em 104},~503--509.
\newblock {\url{https://doi.org/10.1021/jp992980j}}.

\bibitem[Yao \em{et~al.}(2012)Yao, Zheng, Cai, Li, Song, and Zuo]{Yao12}
Yao, M.X.; Zheng, Q.; Cai, X.M.; Li, Y.Z.; Song, Y.; Zuo, J.L.
\newblock Chiral Cyanide-Bridged Cr$^{\rm III}$--Mn$^{\rm III}$
  Heterobimetallic Chains Based on [(Tp)Cr(CN)$_3$]$^{-}$: Synthesis,
  Structures, and Magnetic Properties.
\newblock {\em Inorg. Chem.} {\bf 2012}, {\em 51},~2140--2149.
\newblock PMID: 22303859, {\url{https://doi.org/10.1021/ic201982d}}.

\bibitem[Meng \em{et~al.}(2019)Meng, Shi, and Cheng]{MENG2019134}
Meng, X.; Shi, W.; Cheng, P.
\newblock Magnetism in one-dimensional metal-itronyl nitroxide radical system.
\newblock {\em Coordination Chemistry Reviews} {\bf 2019}, {\em 378},~134--150.
\newblock {\url{https://doi.org/https://doi.org/10.1016/j.ccr.2018.02.002}}.

\bibitem[Thorarinsdottir and Harris(2020)]{Thorarinsdottir20}
Thorarinsdottir, A.E.; Harris, T.D.
\newblock Metal-Organic Framework Magnets.
\newblock {\em Chemical Reviews} {\bf 2020}, {\em 120},~8716--8789.
\newblock PMID: 32045215, {\url{https://doi.org/10.1021/acs.chemrev.9b00666}}.

\bibitem[Yamaguchi \em{et~al.}(2021)Yamaguchi, Okita, Iwasaki, Kono, Hosokoshi,
  Kida, Matsuo, Kawakami, and Hagiwara]{Yamaguchi21}
Yamaguchi, H.; Okita, T.; Iwasaki, Y.; Kono, Y.; Hosokoshi, Y.; Kida, T.;
  Matsuo, A.; Kawakami, T.; Hagiwara, M.
\newblock Magnetic Properties of a Mixed Spin-$(1/2,5/2)$ Chain in
  {(4-Cl-$o$-MePy-V)FeCl$_4$}.
\newblock {\em J. Phys. Soc. Jpn.} {\bf 2021}, {\em 90},~064707.
\newblock {\url{https://doi.org/10.7566/JPSJ.90.064707}}.

\bibitem[Fukushima \em{et~al.}(2004)Fukushima, Honecker, Wessel, and
  Brenig]{fukushima}
Fukushima, N.; Honecker, A.; Wessel, S.; Brenig, W.
\newblock Thermodynamic properties of ferromagnetic mixed-spin chain systems.
\newblock {\em Phys. Rev. B} {\bf 2004}, {\em 69},~174430.
\newblock {\url{https://doi.org/10.1103/PhysRevB.69.174430}}.

\bibitem[Fukushima \em{et~al.}(2005)Fukushima, Honecker, Wessel, Grossjohann,
  and Brenig]{FUKUSHIMA20051409}
Fukushima, N.; Honecker, A.; Wessel, S.; Grossjohann, S.; Brenig, W.
\newblock Specific heat and magnetic susceptibility of ferromagnetic mixed-spin
  chain systems.
\newblock {\em Physica B} {\bf 2005}, {\em 359-361},~1409--1411.
\newblock {\url{https://doi.org/https://doi.org/10.1016/j.physb.2005.01.443}}.

\bibitem[Abouie \em{et~al.}(2006)Abouie, Ghasemi, and
  Langari]{PhysRevB.73.014411}
Abouie, J.; Ghasemi, S.A.; Langari, A.
\newblock Thermodynamic properties of ferrimagnetic spin chains in the presence
  of a magnetic field.
\newblock {\em Phys. Rev. B} {\bf 2006}, {\em 73},~014411.
\newblock {\url{https://doi.org/10.1103/PhysRevB.73.014411}}.

\bibitem[Boyarchenkov \em{et~al.}(2007)Boyarchenkov, Bostrem, and
  Ovchinnikov]{PhysRevB.76.224410}
Boyarchenkov, A.S.; Bostrem, I.G.; Ovchinnikov, A.S.
\newblock Quantum magnetization plateau and sign change of the magnetocaloric
  effect in a ferrimagnetic spin chain.
\newblock {\em Phys. Rev. B} {\bf 2007}, {\em 76},~224410.
\newblock {\url{https://doi.org/10.1103/PhysRevB.76.224410}}.

\bibitem[Yuan \em{et~al.}(2010)Yuan, Ying, and Chuang-Chuang]{Yuan_2010}
Yuan, C.; Ying, X.; Chuang-Chuang, S.
\newblock Magnetic Properties of One-Dimensional Ferromagnetic Mixed-Spin Model
  within {T}yablikov Decoupling Approximation.
\newblock {\em Commun. Theor. Phys.} {\bf 2010}, {\em 54},~747--752.
\newblock {\url{https://doi.org/10.1088/0253-6102/54/4/30}}.

\bibitem[Hu \em{et~al.}(2015)Hu, Wu, Cui, and Qin]{HU2015539}
Hu, A.Y.; Wu, Z.M.; Cui, Y.T.; Qin, G.P.
\newblock The paramagnetic properties of ferromagnetic mixed-spin chain system.
\newblock {\em J. Magn. Magn. Mater.} {\bf 2015}, {\em 374},~539--543.
\newblock {\url{https://doi.org/https://doi.org/10.1016/j.jmmm.2014.09.010}}.

\bibitem[Yan \em{et~al.}(2015)Yan, Zhu, and Su]{Yan15}
Yan, X.; Zhu, Z.G.; Su, G.
\newblock Combined study of Schwinger-boson mean-field theory and linearized
  tensor renormalization group on Heisenberg ferromagnetic mixed spin $(S,
  \sigma)$ chains.
\newblock {\em AIP Advances} {\bf 2015}, {\em 5},~077183.
\newblock {\url{https://doi.org/10.1063/1.4927854}}.

\bibitem[da~Silva and Montenegro-Filho(2021)]{Silva2021}
da~Silva, W.M.; Montenegro-Filho, R.R.
\newblock Role of density-dependent magnon hopping and magnon-magnon repulsion
  in ferrimagnetic spin-(1/2, $S$) chains in a magnetic field.
\newblock {\em Phys. Rev. B} {\bf 2021}, {\em 103},~054432.
\newblock {\url{https://doi.org/10.1103/PhysRevB.103.054432}}.

\bibitem[Takahashi(1999)]{takahashi99}
Takahashi, M.
\newblock {\em Thermodynamics of One-Dimensional Solvable Models}; Cambridge
  University Press,  1999.
\newblock {\url{https://doi.org/10.1017/CBO9780511524332}}.

\bibitem[Dembi\'nski and Wydro(1975)]{Dembiski75}
Dembi\'nski, S.T.; Wydro, T.
\newblock Linear quantum-classical {H}eisenberg model.
\newblock {\em physica status solidi (b)} {\bf 1975}, {\em 67},~K123--K126.
\newblock {\url{https://doi.org/10.1002/pssb.2220670246}}.

\bibitem[Hagiwara \em{et~al.}(1998)Hagiwara, Minami, Narumi, Tatani, and
  Kindo]{Hagiwara1998}
Hagiwara, M.; Minami, K.; Narumi, Y.; Tatani, K.; Kindo, K.
\newblock Magnetic Properties of a Quantum Ferrimagnet:
  {NiCu}(pba)({D}$_2${O})$_3\cdot2${D}$_2${O}.
\newblock {\em J. Phys. Soc. Jpn.} {\bf 1998}, {\em 67},~2209--2211.
\newblock {\url{https://doi.org/10.1143/JPSJ.67.2209}}.

\bibitem[Hagiwara \em{et~al.}(1999)Hagiwara, Narumi, Minami, Tatani, and
  Kindo]{Hagiwara1999}
Hagiwara, M.; Narumi, Y.; Minami, K.; Tatani, K.; Kindo, K.
\newblock Magnetization Process of the {$S= 1/2$} and $1$ Ferrimagnetic Chain
  and Dimer.
\newblock {\em J. Phys. Soc. Jpn.} {\bf 1999}, {\em 68},~2214--2217.
\newblock {\url{https://doi.org/10.1143/JPSJ.68.2214}}.

\bibitem[Fujiwara and Hagiwara(2000)]{Fujiwara2000}
Fujiwara, N.; Hagiwara, M.
\newblock Low energy spin dynamics of a quantum ferrimagnetic chain,
  {NiCu}(pba)({H}$_2${O})$_3$2{H}$_2${O}.
\newblock {\em Solid State Commun.} {\bf 2000}, {\em 113},~433 -- 436.
\newblock
  {\url{https://doi.org/https://doi.org/10.1016/S0038-1098(99)00515-3}}.

\bibitem[Wynn \em{et~al.}(1997)Wynn, G\^{\i}r\c{t}u, Miller, and Epstein]{wynn}
Wynn, C.M.; G\^{\i}r\c{t}u, M.A.; Miller, J.S.; Epstein, A.J.
\newblock Lattice- and spin-dimensionality crossovers in a
  linear-chain-molecule-based ferrimagnet with weak spin anisotropy.
\newblock {\em Phys. Rev. B} {\bf 1997}, {\em 56},~315--320.
\newblock {\url{https://doi.org/10.1103/PhysRevB.56.315}}.

\bibitem[Affronte \em{et~al.}(1999)Affronte, Caneschi, Cucci, Gatteschi,
  Lasjaunias, Paulsen, Pini, Rettori, and Sessoli]{affronte99}
Affronte, M.; Caneschi, A.; Cucci, C.; Gatteschi, D.; Lasjaunias, J.C.;
  Paulsen, C.; Pini, M.G.; Rettori, A.; Sessoli, R.
\newblock Low-temperature thermodynamic properties of molecular magnetic
  chains.
\newblock {\em Phys. Rev. B} {\bf 1999}, {\em 59},~6282--6293.
\newblock {\url{https://doi.org/10.1103/PhysRevB.59.6282}}.

\bibitem[G\^{\i}r\c{t}u \em{et~al.}(2000)G\^{\i}r\c{t}u, Wynn, Zhang, Miller,
  and Epstein]{Girtu00}
G\^{\i}r\c{t}u, M.A.; Wynn, C.M.; Zhang, J.; Miller, J.S.; Epstein, A.J.
\newblock Magnetic properties and critical behavior of
  {Fe}(tetracyanoethylene)$_2 \cdot x$({CH}$_2${Cl}$_2$): A high-${T}_c$
  molecule-based magnet.
\newblock {\em Phys. Rev. B} {\bf 2000}, {\em 61},~492--500.
\newblock {\url{https://doi.org/10.1103/PhysRevB.61.492}}.

\bibitem[Lascialfari \em{et~al.}(2003)Lascialfari, Ullu, Affronte, Cinti,
  Caneschi, Gatteschi, Rovai, Pini, and Rettori]{Lascialfari03}
Lascialfari, A.; Ullu, R.; Affronte, M.; Cinti, F.; Caneschi, A.; Gatteschi,
  D.; Rovai, D.; Pini, M.G.; Rettori, A.
\newblock Specific heat and $\mu^{+}${SR} measurements in
  {$Gd$}(hfac)$_3${NITiPr} molecular magnetic chains: Indications for a chiral
  phase without long-range helical order.
\newblock {\em Phys. Rev. B} {\bf 2003}, {\em 67},~224408.
\newblock {\url{https://doi.org/10.1103/PhysRevB.67.224408}}.

\bibitem[Gillon \em{et~al.}(2002)Gillon, Mathoni\`ere, Ruiz, Alvarez, Cousson,
  Rajendiran, and Kahn]{Gillon2002}
Gillon, B.; Mathoni\`ere, C.; Ruiz, E.; Alvarez, S.; Cousson, A.; Rajendiran,
  T.M.; Kahn, O.
\newblock Spin Densities in a Ferromagnetic Bimetallic Chain Compound:
  Polarized Neutron Diffraction and {DFT} Calculations.
\newblock {\em J. Am. Chem. Soc.} {\bf 2002}, {\em 124},~14433--14441.
\newblock {\url{https://doi.org/10.1021/ja020188h}}.

\bibitem[Kahn \em{et~al.}(1997)Kahn, Bakalbassis, Mathoni\`ere, Hagiwara,
  Katsumata, and Ouahab]{Kahn97}
Kahn, O.; Bakalbassis, E.; Mathoni\`ere, C.; Hagiwara, M.; Katsumata, K.;
  Ouahab, L.
\newblock Metamagnetic Behavior of the Novel Bimetallic Ferromagnetic Chain
  Compound {MnNi}({NO}$_2$)$_4$(en)$_2$ (en = Ethylenediamine).
\newblock {\em Inorg. Chem.} {\bf 1997}, {\em 36},~1530--1531.
\newblock {\url{https://doi.org/10.1021/ic9611453}}.

\bibitem[Feyerherm \em{et~al.}(2001)Feyerherm, Mathoni{\`{e}}re, and
  Kahn]{feyerherm}
Feyerherm, R.; Mathoni{\`{e}}re, C.; Kahn, O.
\newblock Magnetic anisotropy and metamagnetic behaviour of the bimetallic
  chain {MnNi}({NO}$_2$)$_4$(en)$_2$(en = ethylenediamine).
\newblock {\em J. Phys.: Condens. Matter} {\bf 2001}, {\em 13},~2639--2650.
\newblock {\url{https://doi.org/10.1088/0953-8984/13/11/319}}.

\bibitem[Kreitlow \em{et~al.}(2005)Kreitlow, Mathoni\`ere, Feyerherm, and
  S\"ullow]{KREITLOW20052413}
Kreitlow, J.; Mathoni\`ere, C.; Feyerherm, R.; S\"ullow, S.
\newblock Pressure response of the bimetallic chain compound
  MnNi(NO$_2$)$_4$(en)$_2$; en=ethylenediamine.
\newblock {\em Polyhedron} {\bf 2005}, {\em 24},~2413--2416.
\newblock {\url{https://doi.org/https://doi.org/10.1016/j.poly.2005.03.122}}.

\bibitem[Troyer \em{et~al.}(1998)Troyer, Ammon, and Heeb]{alps1}
Troyer, M.; Ammon, B.; Heeb, E.
\newblock Parallel Object Oriented Monte Carlo Simulations.
\newblock {\em Lecture Notes in Computer Science} {\bf 1998}, {\em
  1505},~191--198.
\newblock {\url{https://doi.org/10.1007/3-540-49372-7_20}}.

\bibitem[Albuquerque \em{et~al.}(2007)Albuquerque, Alet, Corboz, Dayal,
  Feiguin, Fuchs, Gamper, Gull, G\"urtler, Honecker, Igarashi, Körner,
  Kozhevnikov, L\"auchli, Manmana, Matsumoto, McCulloch, Michel, Noack,
  Paw{\l}owski, Pollet, Pruschke, Schollw\"ock, Todo, Trebst, Troyer, Werner,
  and Wessel]{alps2}
Albuquerque, A.; Alet, F.; Corboz, P.; Dayal, P.; Feiguin, A.; Fuchs, S.;
  Gamper, L.; Gull, E.; G\"urtler, S.; Honecker, A.;  et~al.
\newblock The {ALPS} project release 1.3: Open-source software for strongly
  correlated systems.
\newblock {\em J. Magn. Magn. Mater.} {\bf 2007}, {\em 310},~1187--1193.
\newblock {\url{https://doi.org/10.1016/j.jmmm.2006.10.304}}.

\bibitem[Alet \em{et~al.}(2005)Alet, Wessel, and Troyer]{alps-sse}
Alet, F.; Wessel, S.; Troyer, M.
\newblock Generalized directed loop method for quantum {M}onte {C}arlo
  simulations.
\newblock {\em Phys. Rev. E} {\bf 2005}, {\em 71},~036706.
\newblock {\url{https://doi.org/10.1103/PhysRevE.71.036706}}.

\bibitem[Todo and Kato(2001)]{PhysRevLett.87.047203}
Todo, S.; Kato, K.
\newblock Cluster Algorithms for General-${S}$ Quantum Spin Systems.
\newblock {\em Phys. Rev. Lett.} {\bf 2001}, {\em 87},~047203.
\newblock {\url{https://doi.org/10.1103/PhysRevLett.87.047203}}.

\bibitem[Sylju\aa{}sen and Sandvik(2002)]{Sandvik}
Sylju\aa{}sen, O.F.; Sandvik, A.W.
\newblock Quantum {M}onte {C}arlo with directed loops.
\newblock {\em Phys. Rev. E} {\bf 2002}, {\em 66},~046701.
\newblock {\url{https://doi.org/10.1103/PhysRevE.66.046701}}.

\bibitem[Bauer \em{et~al.}(2011)Bauer, Carr, Evertz, Feiguin, Freire, Fuchs,
  Gamper, Gukelberger, Gull, Guertler, Hehn, Igarashi, Isakov, Koop, Ma, Mates,
  Matsuo, Parcollet, Paw{\l}owski, Picon, Pollet, Santos, Scarola, Schollwöck,
  Silva, Surer, Todo, Trebst, Troyer, Wall, Werner, and Wessel]{Bauer_2011}
Bauer, B.; Carr, L.D.; Evertz, H.G.; Feiguin, A.; Freire, J.; Fuchs, S.;
  Gamper, L.; Gukelberger, J.; Gull, E.; Guertler, S.;  et~al.
\newblock The {ALPS} project release 2.0: open source software for strongly
  correlated systems.
\newblock {\em J. Stat. Mech.: Theor. Exp.} {\bf 2011}, {\em 2011},~P05001.
\newblock {\url{https://doi.org/10.1088/1742-5468/2011/05/p05001}}.

\bibitem[Matsumoto and Nishimura(1998)]{MTrng}
Matsumoto, M.; Nishimura, T.
\newblock Mersenne Twister: A 623-dimensionally Equidistributed Uniform
  Pseudo-random Number Generator.
\newblock {\em ACM Trans. Model. Comput. Simul.} {\bf 1998}, {\em 8},~3--30.
\newblock {\url{https://doi.org/10.1145/272991.272995}}.

\bibitem[Gvozdikova \em{et~al.}(2016)Gvozdikova, Ziman, and Zhitomirsky]{GZZ16}
Gvozdikova, M.V.; Ziman, T.; Zhitomirsky, M.E.
\newblock Helicity, anisotropies, and their competition in a multiferroic
  magnet: Insight from the phase diagram.
\newblock {\em Phys. Rev. B} {\bf 2016}, {\em 94},~020406.
\newblock {\url{https://doi.org/10.1103/PhysRevB.94.020406}}.

\bibitem[Schulz(1996)]{Schulz}
Schulz, H.J.
\newblock Dynamics of Coupled Quantum Spin Chains.
\newblock {\em Phys. Rev. Lett.} {\bf 1996}, {\em 77},~2790--2793.
\newblock {\url{https://doi.org/10.1103/PhysRevLett.77.2790}}.

\bibitem[Cavadini \em{et~al.}(2000)Cavadini, R\"uegg, Henggeler, Furrer,
  G\"udel, Kr\"amer, and Mutka]{Cavadini2000}
Cavadini, N.; R\"uegg, C.; Henggeler, W.; Furrer, A.; G\"udel, H.U.; Kr\"amer,
  K.; Mutka, H.
\newblock Temperature renormalization of the magnetic excitations in S=1/2
  KCuCl$_3$.
\newblock {\em Eur. Phys. J. B} {\bf 2000}, {\em 18},~565--571.
\newblock {\url{https://doi.org/10.1007/s100510070003}}.

\bibitem[Todo and Shibasaki(2008)]{TodoShibasaki}
Todo, S.; Shibasaki, A.
\newblock Improved chain mean-field theory for quasi-one-dimensional quantum
  magnets.
\newblock {\em Phys. Rev. B} {\bf 2008}, {\em 78},~224411.
\newblock {\url{https://doi.org/10.1103/PhysRevB.78.224411}}.

\bibitem[Fazekas(1999)]{Fazekas}
Fazekas, P.
\newblock {\em Lecture Notes on Electron Correlation and Magnetism}; World
  Scientific: Singapore,  1999.
\newblock {\url{https://doi.org/10.1142/2945}}.

\bibitem[Grossjohann(2004)]{GrDiplom}
Grossjohann, S.
\newblock Stochastic Series Expansion an niedrigdimensionalen
  Quanten-Spin-Systemen.
\newblock Diplomarbeit, TU Braunschweig,  2004.

\bibitem[Trippe \em{et~al.}(2010)Trippe, Honecker, Kl\"umper, and
  Ohanyan]{mceTrippe}
Trippe, C.; Honecker, A.; Kl\"umper, A.; Ohanyan, V.
\newblock Exact calculation of the magnetocaloric effect in the
  spin-$\frac{1}{2}$ {$XXZ$} chain.
\newblock {\em Phys. Rev. B} {\bf 2010}, {\em 81},~054402.
\newblock {\url{https://doi.org/10.1103/PhysRevB.81.054402}}.

\bibitem[Landau(1937)]{Landau37}
Landau, L.D.
\newblock On the theory of phase transitions {I}.
\newblock {\em Zh. Eksp. Teor. Fiz.} {\bf 1937}, {\em 7},~19--32.

\bibitem[Wolf \em{et~al.}(2014)Wolf, Honecker, Hofstetter, Tutsch, and
  Lang]{Wolf14}
Wolf, B.; Honecker, A.; Hofstetter, W.; Tutsch, U.; Lang, M.
\newblock Cooling through quantum criticality and many-body effects in
  condensed matter and cold gases.
\newblock {\em Int. J. Mod. Phys. B} {\bf 2014}, {\em 28},~1430017.
\newblock {\url{https://doi.org/10.1142/S0217979214300175}}.

\bibitem[Konieczny \em{et~al.}(2022)Konieczny, Sas, Czernia, Pacanowska, Fitta,
  and Pe{\l}ka]{Konieczny22}
Konieczny, P.; Sas, W.; Czernia, D.; Pacanowska, A.; Fitta, M.; Pe{\l}ka, R.
\newblock Magnetic cooling: a molecular perspective.
\newblock {\em Dalton Trans.} {\bf 2022}, {\em 51},~12762--12780.
\newblock {\url{https://doi.org/10.1039/D2DT01565J}}.

\bibitem[Junger \em{et~al.}(2005)Junger, Ihle, and Richter]{JIR05}
Junger, I.J.; Ihle, D.; Richter, J.
\newblock Thermodynamics of {$S\ge1$} ferromagnetic {H}eisenberg chains with
  uniaxial single-ion anisotropy.
\newblock {\em Phys. Rev. B} {\bf 2005}, {\em 72},~064454.
\newblock {\url{https://doi.org/10.1103/PhysRevB.72.064454}}.

\bibitem[Tiwari(2022)]{TiwariPhD}
Tiwari, M.
\newblock Mean-field theory for quantum spin systems and the magnetocaloric
  effect.
\newblock {Ph.D.} thesis, CY Cergy Paris Université,  2022.

\bibitem[Weiss(1907)]{weiss07}
Weiss, P.
\newblock L'hypoth\`ese du champ mol\'eculaire et la propri\'et\'e
  ferromagn\'etique.
\newblock {\em J. Phys. Theor. Appl.} {\bf 1907}, {\em 6},~661--690.
\newblock {\url{https://doi.org/10.1051/jphystap:019070060066100}}.

\bibitem[Blundell(2001)]{Blundell01}
Blundell, S.
\newblock {\em Magnetism in Condensed Matter}; Oxford University Press,  2001.

\bibitem[Pires(2021)]{PiresTeixeira21}
Pires, A.S.T.
\newblock The {H}eisenberg model. In {\em Theoretical Tools for Spin Models in
  Magnetic Systems}; IOP Publishing,  2021; pp. 1--1 to 1--16.
\newblock {\url{https://doi.org/10.1088/978-0-7503-3879-0ch1}}.

\bibitem[Melchy and Zhitomirsky(2009)]{Melchy09}
Melchy, P.E.; Zhitomirsky, M.E.
\newblock Interplay of anisotropy and frustration: Triple transitions in a
  triangular-lattice antiferromagnet.
\newblock {\em Phys. Rev. B} {\bf 2009}, {\em 80},~064411.
\newblock {\url{https://doi.org/10.1103/PhysRevB.80.064411}}.

\bibitem[Zhitomirsky and Nikuni(1998)]{ZN98}
Zhitomirsky, M.E.; Nikuni, T.
\newblock Magnetization curve of a square-lattice {H}eisenberg antiferromagnet.
\newblock {\em Phys. Rev. B} {\bf 1998}, {\em 57},~5013--5016.
\newblock {\url{https://doi.org/10.1103/PhysRevB.57.5013}}.

\bibitem[Mermin and Wagner(1966)]{MW66}
Mermin, N.D.; Wagner, H.
\newblock Absence of Ferromagnetism or Antiferromagnetism in One- or
  Two-Dimensional Isotropic Heisenberg Models.
\newblock {\em Phys. Rev. Lett.} {\bf 1966}, {\em 17},~1133--1136.
\newblock {\url{https://doi.org/10.1103/PhysRevLett.17.1133}}.

\bibitem[Fisher(1975)]{Fisher75}
Fisher, M.E.
\newblock Theory of multicritical transitions and the spin-flop bicritical
  point.
\newblock {\em AIP Conf. Proc.} {\bf 1975}, {\em 24},~273--280.
\newblock {\url{https://doi.org/10.1063/1.30084}}.

\bibitem[Landau and Binder(1978)]{LB78}
Landau, D.P.; Binder, K.
\newblock Phase diagrams and multicritical behavior of a three-dimensional
  anisotropic Heisenberg antiferromagnet.
\newblock {\em Phys. Rev. B} {\bf 1978}, {\em 17},~2328--2342.
\newblock {\url{https://doi.org/10.1103/PhysRevB.17.2328}}.

\bibitem[Hassani(1988)]{HassaniMaster}
Hassani, Y.
\newblock Magnetic phase diagram of the two-dimensional {H}eisenberg spin
  one-half canted antiferromagnet ethyl-ammonium tetrabromocuprate({II}).
\newblock Master thesis, Montana State University,  1988.

\bibitem[Selke \em{et~al.}(2009)Selke, Bannasch, Holtschneider, McCulloch,
  Peters, and Wessel]{Selke09}
Selke, W.; Bannasch, G.; Holtschneider, M.; McCulloch, I.P.; Peters, D.;
  Wessel, S.
\newblock Classical and quantum anisotropic Heisenberg antiferromagnets.
\newblock {\em Condensed Matter Physics} {\bf 2009}, {\em 12},~547--558.
\newblock {\url{https://doi.org/10.5488/CMP.12.4.547}}.

\bibitem[Tishin and Spichkin(2003)]{TishinSpichkin}
Tishin, A.M.; Spichkin, Y.I.
\newblock {\em The Magnetocaloric Effect and its Applications}; CRC Press: Boca
  Raton,  2003.
\newblock {\url{https://doi.org/doi.org/10.1201/9781420033373}}.

\bibitem[Heydarinasab and Abouie(2020)]{Heydarinasab2020}
Heydarinasab, F.; Abouie, J.
\newblock Mixed-spin system with supersolid phases: magnetocaloric effect and
  thermal properties.
\newblock {\em J. Phys.: Condens. Matter} {\bf 2020}, {\em 32},~165804.
\newblock {\url{https://doi.org/10.1088/1361-648x/ab61ca}}.

\bibitem[Palacios \em{et~al.}(2022)Palacios, S\'aez-Puche, Romero, Doi,
  Hinatsu, and Evangelisti]{Palacios2022}
Palacios, E.; S\'aez-Puche, R.; Romero, J.; Doi, Y.; Hinatsu, Y.; Evangelisti,
  M.
\newblock Large magnetocaloric effect in {EuGd$_2$O$_4$} and {EuDy$_2$O$_4$}.
\newblock {\em J. Alloy. Compd.} {\bf 2022}, {\em 890},~161847.
\newblock
  {\url{https://doi.org/https://doi.org/10.1016/j.jallcom.2021.161847}}.

\end{thebibliography}

\end{adjustwidth}
\end{document}